\documentclass[showpacs,aps,prd,preprint,superscriptaddress,nofootinbib,tightenlines]{revtex4-1}
\usepackage[colorlinks=true, pdfstartview=FitV, linkcolor=red,citecolor=blue, urlcolor=blue,
bookmarks=true, bookmarksnumbered=true, breaklinks]{hyperref}
\usepackage[dvipdfmx]{graphicx}
\usepackage{amsmath,amssymb,bm,ascmac,color,longtable,mathrsfs}

\usepackage{slashed}

\allowdisplaybreaks[1]

\usepackage{color}

\usepackage[normalem]{ulem}

\def\simge{\mathrel{
   \rlap{\raise 0.511ex \hbox{$>$}}{\lower 0.511ex \hbox{$\sim$}}}}
\def\simle{\mathrel{
   \rlap{\raise 0.511ex \hbox{$<$}}{\lower 0.511ex \hbox{$\sim$}}}}
\def\bigs{\mathrel{
   \rlap{\raise 0.531ex \hbox{$>$}}{\lower 0.531ex \hbox{$<$}}}}

\allowdisplaybreaks[3]
\allowdisplaybreaks

\begin{document}

\title{Kondo phase diagram of quark matter}

\author{Shigehiro~Yasui}
\email[]{yasuis@th.phys.titech.ac.jp}
\affiliation{Department of Physics, Tokyo Institute of Technology, Tokyo 152-8551, Japan}
\author{Kei~Suzuki}
\affiliation{Department of Physics, Tokyo Institute of Technology, Tokyo 152-8551, Japan}
\author{Kazunori~Itakura}
\affiliation{KEK Theory Center, Institute of Particle and Nuclear
Studies, High Energy Accelerator Research Organization, 1-1, Oho,
Ibaraki, 305-0801, Japan}
\affiliation{Graduate University for Advanced Studies (SOKENDAI),
1-1 Oho, Tsukuba, Ibaraki 305-0801, Japan}

\begin{abstract}
We find a novel type of phase, ``the Kondo phase," in a quark matter containing heavy quarks as impurities. This phase allows the QCD Kondo effect to occur between a light quark ($\psi$) and a heavy quark ($\Psi$), and is characterized by a nonzero heavy-light condensate $\langle \bar \psi \Psi\rangle $ which measures the strength of mixing between $\psi$ and $\Psi$ and represents formation of the Kondo cloud, the ground state composed of light and heavy quarks. Within a field-theoretic approach and mean-field approximation, we obtain the gap equation which self-consistently and non-perturbatively determines the gap $\Delta \sim \langle \bar \psi \Psi\rangle $. By solving the gap equation with the condition for the heavy-quark number conservation, we draw a phase diagram in $\mu$ (the light-quark chemical potential) and $\lambda$ (an analog of the heavy-quark chemical potential) plane, and identify the density region of the Kondo phase. 
\end{abstract}


\maketitle

%\tableofcontents

\section{Introduction}
\label{sec:introduction}
The Kondo effect constitutes one of the most fundamental concepts in quantum physics. It originally refers to the increasing behavior of the resistance of a metal containing magnetic impurities with decreasing temperature and is now understood as resulting from the enhanced non-Abelian (spin-exchange) interaction between the conducting electron and the impurity as first revealed by J. Kondo~\cite{Kondo:1964}. The Kondo effect occurs for any small coupling, and leads to formation of the ``Kondo cloud" -- a composite state made of conducting electrons and the impurity as the ground state~\cite{Hewson,Yosida,Yamada}. 
The concept behind the Kondo effect can be universally applied to various systems from metals to artificial materials such as quantum dots, quantum devices~\cite{Goldhaber-Gordon:1998,Cronenwett:1998,Wiel:2000,Jeong:2001,Park:2002,Hur:2015,Balatsky:2006} and ultracold atomic gases~\cite{PhysRevLett.111.135301}.
Moreover, it has been recently recognized~\cite{Yasui:2013xr,Hattori:2015hka,Ozaki:2015sya} that a quark matter containing heavy (charm and bottom) quarks as impurities satisfies the necessary conditions for the Kondo effect to occur~\cite{Hewson,Yosida,Yamada}~\footnote{In addition to the existence of heavy impurities, three conditions are necessary: Fermi surface of light fermions, non-Abelian interactions between a light fermion and an heavy impurity, and quantum effects corresponding to higher loops in the perturbative analysis.}, and exhibits what we call the QCD Kondo effect.
The QCD Kondo effect would affect various properties of the quark matter through the changes of thermodynamic properties (e.g. heat capacity, susceptibility) as well as of transport properties (e.g. electric/color-electric conductivity). 

Quark matters with heavy quark impurities will be realized in extremely high-density environments such as relativistic heavy-ion collisions at FAIR and J-PARC and the core of neutron stars. Small number of heavy quarks will be provided by initial hard scatterings in the heavy-ion collisions, and by high energy neutrinos as cosmic rays that will transform light quarks into heavy quarks while traversing the compact stars. In both cases, heavy impurities will be randomly distributed in a quark matter and far from chemical equilibrium.
With a random distribution, we may be able to define locally a spatial region where the number of heavy quarks (or heavy antiquarks) is larger than the other places. Similar to the development of cosmological density perturbation due to gravitational attractive force, we expect that  heavy quarks will gather via attractive one-gluon exchange interaction to form a denser state like a droplet. In this droplet, the density of heavy quarks will be relatively high and can be approximately treated as uniform. We are going to investigate the QCD Kondo effect in such a droplet of heavy impurities.

Keeping this situation in mind,
we consider a finite density of heavy quark impurities which are uniformly distributed inside
the quark matter. 
This can be understood as an approximate situation for the droplet.
 To research the properties of the ground state of such a system (the Kondo cloud), we will develop a field theoretic approach that allows for a non-perturbative treatment of the Kondo effect. 
In particular, we will introduce a new type of quark condensate, 
a {\it heavy-light} condensate $\langle \bar{\psi}\Psi \rangle$ for a light quark $\psi$ and a heavy quark $\Psi$, which is exploited for the first time in this work. 
The heavy-light condensate for the QCD Kondo effect is highly contrasted with the other types of quark condensate; 
the light quark-antiquark condensate $\langle \bar{\psi}\psi \rangle$ for 
chiral symmetry breaking~\cite{Nambu:1961tp} and the light quark-quark condensate $\langle \psi\psi \rangle$ for 
color superconductivity~\cite{Fukushima:2010bq,Fukushima:2013rx,Alford:2007xm}.
We know that 
these
two condensates are associated with unstable fluctuations in the quark-antiquark and quark-quark channels, respectively.
Hence we understand it reasonable to introduce the heavy-light condensate that is associated with the infrared instability around the Fermi surface leading to the QCD Kondo effect. 
Having defined the heavy-light condensate, we are able to investigate the QCD Kondo effect non-perturbatively within the mean-field approximation.  
This also implies that the heavy-light condensate would be formed under the same conditions for the Kondo effect to occur.
Indeed, such analysis is successful in the investigation of the Kondo cloud in condensed matter~\cite{Takano:1966,Yoshimori:1970} (see also for the Kondo lattice~\cite{Lacroix:1979,PhysRevB.28.5255,PhysRevB.30.3841}, the quantum dots~\cite{Eto:2001} and Dirac fermions~\cite{Yanagisawa:2015conf,Yanagisawa:2015}). 
Importantly, the mean-field approach can be justified for large $N$ in the $\mathrm{SU}(N)$ symmetry for the non-Abelian interaction~\cite{PhysRevB.28.5255,PhysRevB.30.3841}.
Furthermore, the isospin Kondo effect in ``charm (bottom) nuclei" was recently studied in the mean-field approach by one of the authors~\cite{Yasui:2016ngy}~\footnote{See Ref.~\cite{Sugawara-Tanabe:1979} for early application of the Kondo effect in deformed nuclei.}.

In this work, by using a simple model that exhibits the QCD Kondo effect, we will determine the ground state of the quark matter with heavy impurities with light and heavy quark densities being varied, and will identify the ``Kondo phase" in which the heavy-light condensate $\langle \bar{\psi}\Psi \rangle$ is finite and thus the QCD Kondo effect is at work. Although the results in this analysis are obtained with heavy quarks uniformly distributed in the whole space,  one can translate them into those of a droplet through the ``energy gain" per a single heavy quark.

\section{Formalism}
\label{sec:formalism}

\subsection{Lagrangian}
\label{sec:lagrangian}
We consider a model having the current-current interaction with the color exchange between a light (massless) quark ($\psi$) and a heavy quark ($\Psi$) with mass $m_{Q}$~\cite{Klimt:1989pm,Vogl:1989ea,Klevansky:1992qe,Hatsuda:1994pi}: 
\begin{eqnarray}
{\cal L} 
= \bar{\psi} i \partial\hspace{-0.55em}/ \psi + \mu \, \bar{\psi}\gamma_{0}\psi + \bar{\Psi} i \partial\hspace{-0.55em}/ \Psi - m_{Q} \bar{\Psi}\Psi
- G_{c} \sum_{a} \left( \bar{\psi} \gamma^{\mu} T^{a} \psi \right)
                       \left( \bar{\Psi} \gamma_{\mu} T^{a} \Psi \right)\, ,
\label{eq:L}
\end{eqnarray}
where  $T^{a}=\lambda^{a}/2$ ($a=1,\dots,N_c^2-1$) are the generators of SU($N_c$) with $\lambda^{a}$ being the Gell-Mann matrices and $N_{c}=3$ in QCD.
The coupling strength $G_{c}$ is taken to be positive $G_{c}>0$ so that the interaction produces attraction in the color anti-triplet channel mimicking the one-gluon-exchange interaction. 
We consider $N_{f}$ flavors for light quarks: $\psi=(\psi^{1},\dots,\psi^{N_{f}})^{\mathrm{t}}$. For simplicity, flavor indices are not explicitly written for the fields $\psi^{i}$ ($i=1,\dots,N_{f}$) and we assume that they have the common chemical potential $\mu$.
We note that ${\cal L}$ is invariant under both the chiral symmetry for light quarks and the heavy-quark spin symmetry for heavy quarks.

We can simplify the Lagrangian (\ref{eq:L}) in the heavy quark limit according to the standard procedure in the heavy-quark effective theory~\cite{Neubert:1993mb,Manohar:2000dt}: 
We separate the heavy quark momentum $p^{\mu}$ into the on-mass-shell part $m_Qv^\mu$ and the off-mass-shell part $k^\mu$  as $p^{\mu}=m_{Q}v^{\mu}+k^{\mu}$ with $v^{\mu}$ being the four-velocity of the heavy quark ($v^{\mu}v_{\mu}=1$). The energy scale of $k^{\mu}$ should be much smaller than $m_{Q}$. 
We focus on the dynamics concerning the residual momentum $k^{\mu}$. 
Namely, we replace $\Psi$ by the effective field $\Psi_{v}$ having 
the momentum $k^{\mu}$ as 
$\Psi \rightarrow \Psi_{v} = \frac{1}{2}(1+v\hspace{-0.5em}/) {\rm e}^{im_{Q}v\cdot x}\Psi$. The factor $\frac12 (1+v\hspace{-0.5em}/)$ is a projection onto the positive energy components while the factor ${\rm e}^{im_{Q}v\cdot x}$ works to subtract the on-shell motion from a heavy-quark field $\Psi$ so that the effective field $\Psi_v$ contains only the residual motion.
Concerning the four-point interaction, we use the Fierz transformation,
\begin{eqnarray}
\sum_{a} (T^{a})_{ij} (T^{a})_{kl} = \frac{1}{2}\, \delta_{il}\delta_{kj} 
- \frac{1}{2N_{c}}\, \delta_{ij} \delta_{kl}\, , \nonumber 
\end{eqnarray}
for the color part. It is important to use the Fierz transformation only for the SU($N_{c}$) generator, not for Dirac matrices. 
Below, we consider the static limit for the impurity motion $v^{\mu}=(1,\vec{0}\,)$ 
and use the relation $\bar{\Psi}_{v}\Psi_{v} = \Psi_{v}^{\dag}\Psi_{v}$.

Let us define a ``heavy-quark number density" $\Psi_v^\dagger \Psi_v$ from the effective field $\Psi_v$.
Then, we consider the condition, 
$\Psi_{v}^{\dag}(x)\Psi_{v}(x)
=
\sum_{i} \delta^{(3)}(\vec{x}-\vec{x}_{i}) 
\rightarrow
n_{Q} \equiv 
\overline{\sum}_{i} \delta^{(3)}(\vec{x}-\vec{x}_{i})$, 
with the three-dimensional delta function $\delta^{(3)}(\vec{x})$ and $\vec{x}_{i}$ being the position of the impurity $i$.
$\overline{\sum}_{i}$ indicates average over the whole space.
Thus, $n_{Q}$ is the averaged number density of impurities. 
When we consider the impurity distribution in space, we tend to suppose that a single heavy quark exists like a single point in space.
In contrast, for the present analysis, we suppose that the heavy quarks are distributed {\it uniformly} in space, 
and the density is sufficiently large so that the averaged distance between heavy quarks is smaller than the coherence length for the QCD Kondo effect $\simeq 1/|\Delta|$, which will be explained later. 
Due to the uniformity of the heavy quark distribution, the ground state does not break the translational invariance in contrast to the case with a single heavy quark impurity. 
In addition, as mentioned before, we assume that all the heavy quarks are static and do not propagate in space.

We make a comment on the number density of heavy quarks $\Psi_v^\dagger \Psi_v$ which was defined above. This is not equivalent to the number density of heavy quarks in a usual sense because it is defined from the effective heavy quark field $\Psi_v$. In the heavy-quark effective theory, the energy of the original heavy quark is separated into the on-shell component (rest mass $m_{Q}$) and the off-shell component ($k^{0}$) as represented in the momentum decomposition $p^{\mu}=m_{Q}v^{\mu}+k^{\mu}$. Notice that the scattering between a light quark and a heavy quark is mediated by the exchange of a momentum $k^\mu$, while keeping the on-shell component $m_Qv^\mu$ intact, and thus the on-shell component of a heavy quark is irrelevant to the present problem.
Therefore, what matters in the problem 
is the number density of the heavy quarks with non-zero off-shell components ($k^{0}\neq 0$), which is defined by $\Psi_v^\dagger \Psi_v$.

\subsection{Mean-field approximation}
\label{sec:MF}
It has been known within the perturbative renormalization group analysis~\cite{Yasui:2013xr} that the model (\ref{eq:L}) exhibits the QCD Kondo effect for a single impurity case. We treat here the same model in a non-perturbative way when the impurities are homogeneously distributed.
As commented before, we consider the color-singlet correlation between a light quark with 
 flavor $i$ and a heavy impurity $\langle\bar{\psi}^{i}_{\alpha}\Psi_{v\delta}\rangle$ (and its conjugate) which is associated with the instabilities in the QCD Kondo effect. 
Notice that the mean-fields are matrices with Dirac indices $\alpha$ and $\delta$.
Then we perform the mean-field approximation in the Lagrangian (\ref{eq:L})
(represented with respect to the effective heavy impurity field $\Psi_v$):  
\begin{eqnarray}
 (\bar{\psi}^{i}_{\alpha}\Psi_{v\delta})(\bar{\Psi}_{v\gamma}\psi^{i}_{\beta})
\rightarrow
\langle\bar{\psi}^{i}_{\alpha}\Psi_{v\delta}\rangle \bar{\Psi}_{v\gamma}\psi^{i}_{\beta}
+
\langle \bar{\Psi}_{v\gamma}\psi^{i}_{\beta}\rangle \bar{\psi}^{i}_{\alpha}\Psi_{v\delta} 
 -
\langle\bar{\psi}^{i}_{\alpha}\Psi_{v\delta}\rangle \langle \bar{\Psi}_{v\gamma}\psi^{i}_{\beta} \rangle,
\end{eqnarray}
where we have neglected the term of the second order with respect to fluctuation. 
The expectation value, $\langle\bar{\psi}^{i}_{\alpha}\Psi_{v\delta}\rangle$ (or $\langle \bar{\Psi}_{v\gamma}\psi^{i}_{\beta} \rangle$), is evaluated by using the ground state wave function.

Physically, the mean field $\langle\bar{\psi}^{i}_{\alpha}\Psi_{v\delta}\rangle$ should be directly related to the formation of a bound state of a light quark ``hole" and a heavy quark. 
This is analogous to the formation of a $\sigma$ meson in association with the quark-antiquark ($\bar{q}q$) condensate in vacuum~\cite{Klimt:1989pm,Vogl:1989ea,Klevansky:1992qe,Hatsuda:1994pi}.

Let us define the gap function as 
\begin{eqnarray}
\Delta_{\delta \alpha}^{i} &\equiv & \frac{G_{c}}{2} \langle\bar{\psi}^{i}_{\alpha}\Psi_{v\delta}\rangle, 
 \label{eq:gap1} 
\end{eqnarray}
whose Dirac structure is further parametrized in momentum space as
\begin{eqnarray}
\Delta^{i}_{\delta\alpha}=
\Delta^{i} \left( \frac{1+\gamma_{0}}{2} ( 1-\hat{k}\!\cdot\!\vec{\gamma} ) \right)_{\delta\alpha},
\end{eqnarray}
with a scalar (complex) parameter $\Delta^{i}$ and $\hat{k}=\vec{k}/|\vec{k}|$. 
Notice that $\Delta^{i}$ is independent of momentum, which reflects the translational invariance of the ground state. 
Plugging this form of the gap, we finally find the mean-field Lagrangian in the momentum space ($v^\mu=(1,\vec 0)$):
\begin{eqnarray}
{\cal L}^{\mathrm{MF}}
&\!=& \bar{\psi} (k\hspace{-0.45em}/ + \mu \, \gamma_{0} ) \psi + \bar{\Psi}_{v}  v\!\cdot\!k \Psi_{v} 
 -\lambda \left( \Psi^{\dag}_{v}\Psi_{v} - n_{Q} \right) \nonumber \\
&&\hspace{-4mm}
+\! \sum_{i}\! \Delta^{i} \bar{\Psi}_{v} \frac{1+\gamma_{0}}{2} \left( 1+\hat{k}\!\cdot\!\vec{\gamma} \right) \psi^{i} 
+\! \sum_{i}\! \Delta^{i\ast} \bar{\psi}^{i}\! \left( 1+\hat{k}\!\cdot\!\vec{\gamma} \right)\! \frac{1+\gamma_{0}}{2} \Psi_{v} 
-\!\sum_{i}\!\frac{8}{G_{c}} |\Delta^{i}|^{2}\, , 
\end{eqnarray}
where we have added the term with the Lagrange multiplier $\lambda$ to include the constraint 
on the number conservation of the heavy quarks. 
This constraint is necessary because the heavy quark number density should be conserved on average in the mean field (\ref{eq:gap1})
(see Refs.~\cite{Takano:1966,Yoshimori:1970,Eto:2001}). 
Note that $-\lambda$ can be regarded as the energy cost (measured from the heavy quark mass $m_Q$) for putting a heavy quark into a quark matter, and hence it is mathematically equivalent to the chemical potential~\footnote{One may notice that, in case of $\Delta=0$, the number density of heavy quarks $n_{Q}$ jumps discontinuously from zero (for $\lambda>0$) to infinity (for $\lambda<0$). It means that the number density of heavy quarks cannot be uniquely determined by $\lambda$ for $\Delta=0$. This discontinuous behavior stems from the simple setting that the finite mass of heavy quarks is neglected.}.
Also, reflecting the uniformity of the heavy-quark density, the $\lambda$ is taken as constant in space.
However, we should comment that the present formalism can be directly applied to the case where the heavy quark number density is non-uniform in space, $n_Q(\vec x)$.
We also emphasize that the heavy quarks are treated as impurity particles embedded in a quark matter of light flavors.
This also implies that we are going to impose neither charge neutrality or chemical equilibrium with respect to weak interaction. Thus we can regard the $\lambda$  as a parameter independent of the light-quark chemical potential $\mu$.

\begin{figure}[tb]
\begin{center}
  \begin{minipage}[b]{0.32\linewidth}
    \centering
    \includegraphics[keepaspectratio, scale=0.25]{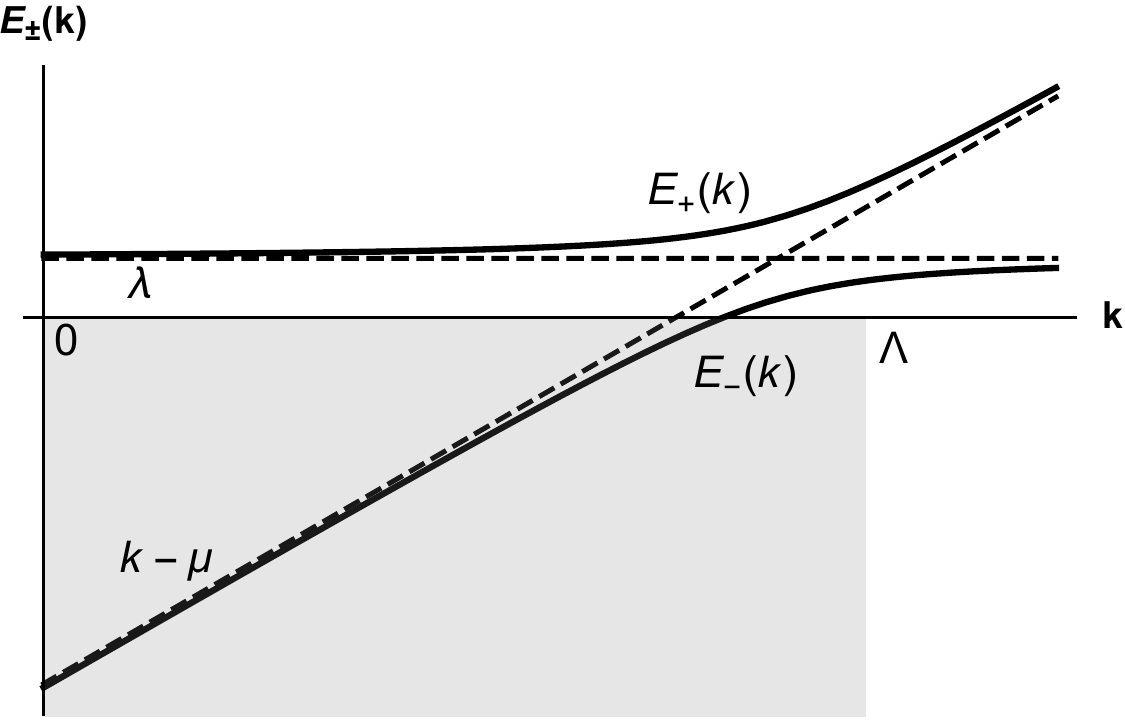}
  \end{minipage}
  \begin{minipage}[b]{0.32\linewidth}
    \centering
    \includegraphics[keepaspectratio, scale=0.25]{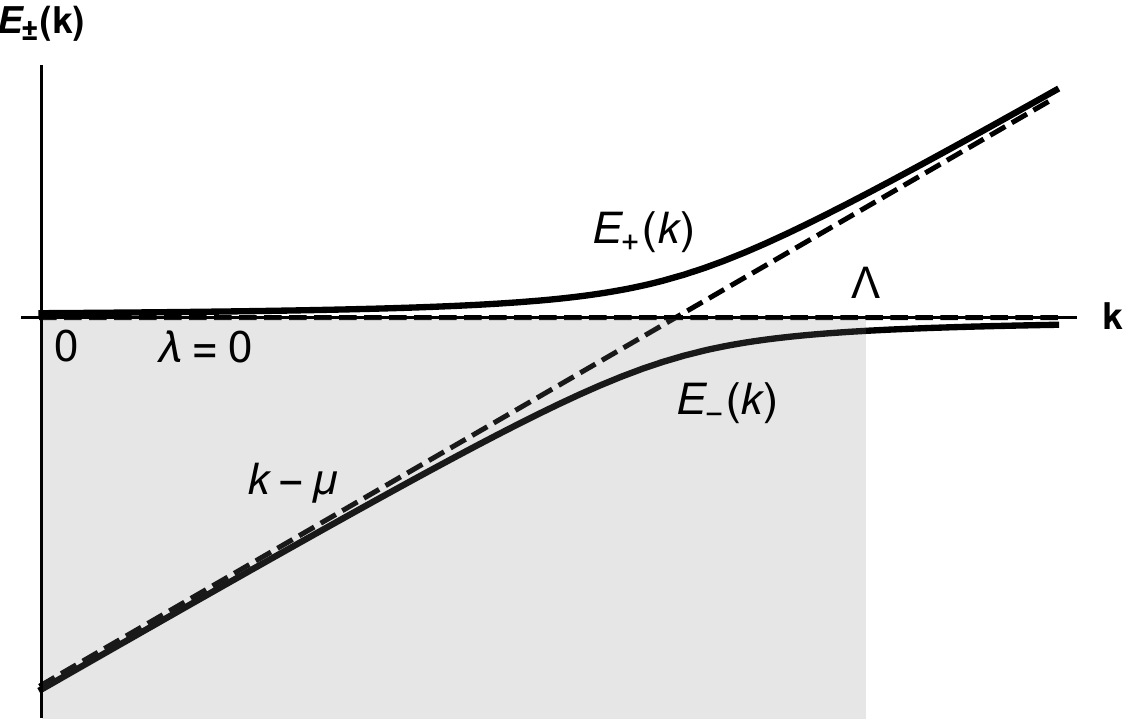}
  \end{minipage}
  \begin{minipage}[b]{0.32\linewidth}
    \centering
    \includegraphics[keepaspectratio, scale=0.25]{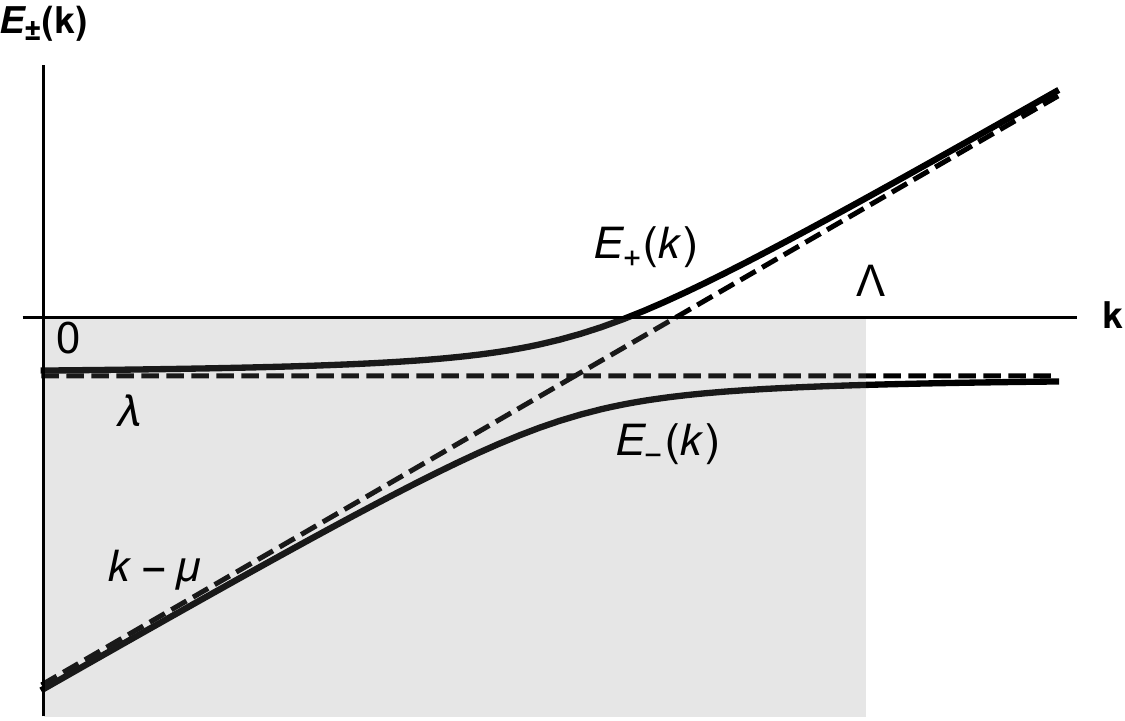}
  \end{minipage}
  \caption{Dispersion relations of quark $E_{\pm}(k)$ with finite gap for positive (left), zero (middle) and negative (right) values of $\lambda$.
   The gray band indicates the region of the integrals, $0<k<\Lambda$.}
  \label{fig:dispersion}
  \end{center}\vspace{-5mm}
\end{figure}

Since the mean-field Lagrangian allows for mixing between the fields $\psi$ and $\Psi_{v}$, we diagonalize it by the Bogoliubov-like transformation to find the following energy-momentum dispersion relations for spin up 
\begin{eqnarray}
\hspace{-4mm}E_{\pm}(k) &\equiv& \frac{1}{2} \left( k\!+\!\lambda\!-\!\mu \!\pm\! \sqrt{(k\!-\!\lambda\!-\!\mu)^{2}\!+\!8N_{f}|\Delta|^{2}} \right),
\label{eq:eigenmode1} \\
E(k) &\equiv& E_i = k - \mu\, \hspace{1.8em} [i=1, \cdots, N_f-1] ,
\label{eq:eigenmode1.5} \\
\tilde{E}(k) &\equiv& \tilde{E}_i = - k - \mu\, \hspace{1.0em} [i=1, \cdots, N_f] ,
\label{eq:eigenmode2}
\end{eqnarray}
where 
 $k=|\vec{k}\,|$ 
 and 
 numbers in the brackets indicate the number of degeneracy.
Here we assume $\Delta^{i}=\Delta$ from light flavor symmetry.
We obtain the same result for spin down.
The relation (\ref{eq:eigenmode1})
is given by 
 the linear combination of $\psi$ and $\Psi_{v}$.
Notice that the mixing takes place only between a positive energy light quark and a heavy quark impurity.
On the other hand, as seen in Eq.~(\ref{eq:eigenmode2}), the light antiquarks are completely decoupled from the heavy quark, because the heavy quark in the effective Lagrangian has no negative-energy component.
The dispersions should possess information about the properties of the ground state in the single-particle picture in the mean-field approximation.
In Fig.~\ref{fig:dispersion} we show the schematic picture of the dispersions~(\ref{eq:eigenmode1}) for positive (left) and negative (right) values of $\lambda$. 
Because we consider high density states, we neglect the negative-energy component (\ref{eq:eigenmode2}).
By using Eqs.~(\ref{eq:eigenmode1}) and (\ref{eq:eigenmode1.5}), we analyze the thermodynamic potential of the ground state.

It is important to note that the three-dimensional momentum of a light quark, $k$, is a conserved quantity because the present 
mean fields maintain the translational invariance of the ground state. 
This is the case as long as the uniform density distribution of the heavy quark is considered. 

We comment that the nonzero value of the gap $\Delta^{i}$ in Eq.~(\ref{eq:gap1}) in the mean-field approximation 
breaks the symmetry in the original Lagrangian, i.e., the chiral symmetry for the light quarks and the heavy-quark (spin) symmetry (HQS) for the heavy quark impurities. We summarize the breaking pattern discussed in detail in Ref.~\cite{Yasui:2017izi}.
The whole symmetry of the system and its breaking pattern is given by
$ G \rightarrow H $
with the group\footnote{The $\mathrm{U}(1)_{\mathrm{A}}$ symmetry exists in our Lagrangian (\ref{eq:L}) although it should be anomalously broken when the fermions couple to gauge fields. If there was no $\mathrm{U}(1)_{\mathrm{A}}$ symmetry, the counterpart $\mathrm{U}(1)_{\mathrm{A}+\mathrm{HQS}_{h}}$ in $H$ would not hold. 
%We note that $\mathrm{U}(1)_{\mathrm{A}}$ holds in the Lagrangian, although this symmetry should already be broken by anomaly effect. 
See also the discussion in the end of section~\ref{sec:results}.}
\begin{eqnarray}
 G
 =
 \mathrm{SO}(3)_{\mathrm{space}} \times \mathrm{SU}(2)_{\mathrm{HQS}} \times \mathrm{U}(1)_{\mathrm{Q}} \times \mathrm{U}(1)_{\mathrm{V}} \times \mathrm{U}(1)_{\mathrm{A}} \times \mathrm{SU}(N_{f})_{\mathrm{R}+\mathrm{L}} \times \mathrm{SU}(N_{f})_{\mathrm{R}-\mathrm{L}},
\label{eq:group_G}
\end{eqnarray}
and the subgroup
\begin{eqnarray}
 H
 =
 \mathrm{SO}(3)_{\mathrm{space}} \times \mathrm{U}(1)_{\mathrm{Q}+\mathrm{V}} \times \mathrm{U}(1)_{\mathrm{A}+\mathrm{HQS}_{h}}
  \times \mathrm{SU}(N_{f}-1)_{\mathrm{R}+\mathrm{L}} \times \mathrm{SU}(N_{f}-1)_{\mathrm{R}-\mathrm{L}}.
\label{eq:group_H}
\end{eqnarray}
In the group $G$, $\mathrm{SO}(3)_{\mathrm{space}}$ is the rotational symmetry in the three-dimensional space. 
The rotational symmetry should be preserved when $n_{Q}$ is constant in space.
We notice that, as more general situations, when $n_{Q}(\vec{x})$ depends on the position $\vec{x}$, this symmetry does not necessarily hold. 
$\mathrm{SU}(2)_{\mathrm{HQS}}$ and  $\mathrm{U}(1)_{\mathrm{Q}}$ are the HQS of $\Psi_{v}$ (spin rotation of the heavy quark) and the vector symmetry for the overall phase of $\Psi_{v}$, respectively.
$\mathrm{U}(1)_{\mathrm{V}} \times \mathrm{U}(1)_{\mathrm{A}} \times \mathrm{SU}(N_{f})_{\mathrm{R}+\mathrm{L}=\mathrm{V}} \times \mathrm{SU}(N_{f})_{\mathrm{R}-\mathrm{L}=\mathrm{A}}$ is the usual chiral symmetry of light flavor $\psi$.
In the subgroup $H$, several combined symmetries appear as explained below.
$\mathrm{U}(1)_{\mathrm{Q}+\mathrm{V}}$ is the simultaneous transformation for $\mathrm{U}(1)_{\mathrm{V}}$ in the light quark and $\mathrm{U}(1)_{\mathrm{Q}}$ in the heavy quark.
$\mathrm{U} (1)_{\mathrm{A}+\mathrm{HQS}_{h}}$ is the symmetry combined by $\mathrm{U}(1)_{\mathrm{A}}$ and $\mathrm{U}(1)_{\mathrm{HQS}_{h}}$.
This is called the {\it chiral-HQS locked ($\chi$HQSL) symmetry}.
We introduce the $\mathrm{U}(1)_{\mathrm{HQS}_{h}}$ symmetry for the U(1) transformation for $\Psi_v$, which is defined by $e^{i\vec\theta\cdot \vec \sigma} \in \mathrm{U}(1) \subset \mathrm{SU}(2)$ with a parameter $\vec\theta=\alpha \hat{p}$ as a subgroup of $\mathrm{SU}(2)_{\mathrm{HQS}}$. 
This transformation conserves the helicity since it is the rotation around the axis along the direction $\hat{p}$ in momentum space.
We comment that it is the factor of $\hat{p} \cdot \vec{\sigma}$ coming from the hedgehog ansatz, $\vec{\Delta}=\Delta \,\hat{p}$, that makes the $\chi$HQSL symmetry possible.
Lastly, $\mathrm{SU}(N_{f}-1)_{\mathrm{R}+\mathrm{L}} \times \mathrm{SU}(N_{f}-1)_{\mathrm{R}-\mathrm{L}}$ means 
that all the light flavors coherently contribute to the formation of gap~\footnote{In this work, all the light flavors are coupled with heavy quarks, so that the chiral symmetry for light flavors is broken due to heavy-light condensates. The residual symmetry $\mathrm{SU}(N_{f}-1)_{\mathrm{R}+\mathrm{L}} \times \mathrm{SU}(N_{f}-1)_{\mathrm{R}-\mathrm{L}}$ in the subgroup $H$ is not the chiral symmetry in the original flavor space ($u,d,s,...$), but the chiral symmetry for a transformed flavor space which is composed of the original flavors. As another choice, if we consider only one light flavor coupled with a heavy quark as Refs.~\cite{Yasui:2017izi,Suzuki:2017gde}, we find the residual chiral symmetry $\mathrm{SU}(N_{f}-1)_{\mathrm{R}+\mathrm{L}} \times \mathrm{SU}(N_{f}-1)_{\mathrm{R}-\mathrm{L}}$ for the decoupled light flavors in the original flavor space.}.

In the numerical calculation, we adopt the three dimensional momentum cutoff because the Lorentz symmetry is violated at finite density. 
We use the parameter values from the usual Nambu--Jona-Lasinio (NJL) model for $N_{f}=2$: $G_{c}=(9/2) \cdot 2.0/\Lambda^{2}$ and $\Lambda=0.65$ GeV.
They are determined 
so as to reproduce the quark condensate and the pion decay constant in vacuum~\cite{Klevansky:1992qe,Hatsuda:1994pi}.
We assume that the interaction between a light quark and a heavy quark has the same strength of the coupling.

\subsection{Thermodynamic potential}
\label{sec:th_pot}
Thermodynamic potential computed from the dispersion relations~(\ref{eq:eigenmode1}) and (\ref{eq:eigenmode1.5}) reads 
\begin{eqnarray}
\Omega(T,\mu,\lambda;\Delta)
=
2N_{c}\!\! \int_{0}^{\Lambda}%\!\!
\frac{k^{2}\mathrm{d}k}{2\pi^2}f(T,\mu,\lambda;k) 
\, +\,  \frac{8N_{f}}{G_{c}} |\Delta|^{2}
\, -\, \lambda \, n_{Q}\, ,\label{eq:thermo_pot}
\end{eqnarray}
with $f(T,\mu,\lambda;k)\equiv -{\beta}^{-1} \ln [( 1+{\rm e}^{ -\beta E_{+}(k)})( 1+{\rm e}^{ -\beta E_{-}(k)})( 1+{\rm e}^{ -\beta E(k)})^{N_{f}-1} ]$ and $\beta=1/T$ being the inverse temperature.
The factor two in the coefficient of the integral comes from the sum of 
 right- and left-handed light quarks.
We introduce $\Lambda$ to make the integral finite.

The value of $|\Delta|$ is determined by the minimum of $\Omega(T,\mu,\lambda;\Delta)$ or the solution to the gap equation:
 $\frac{\partial}{\partial \Delta^{\ast}} \Omega(T,\mu,\lambda;\Delta) = 0$.
The chemical potential $\mu$ and the Lagrange multiplier $\lambda$ are related to the number densities of light quarks and heavy quarks, $n_{q}$ and $n_{Q}$, respectively:
 $-\frac{\partial}{\partial \mu} \Omega(T,\mu,\lambda;\Delta) = n_{q}$
and
$\frac{\partial}{\partial \lambda} \Omega(T,\mu,\lambda;\Delta) = 0$.
The analysis is analogous to that in the color superconductivity with the NJL-type model~\cite{Alford:2007xm,Buballa:2003qv}.
We notice that the light quarks and the heavy quarks are treated on the basis of the grand canonical ensemble. This is essentially important because the number densities for the light quarks and the heavy quarks are not conserved quantities due to the mixing between them by the finite gap $\Delta$.

\begin{figure}[t]
\begin{center}
\includegraphics[scale=0.45]{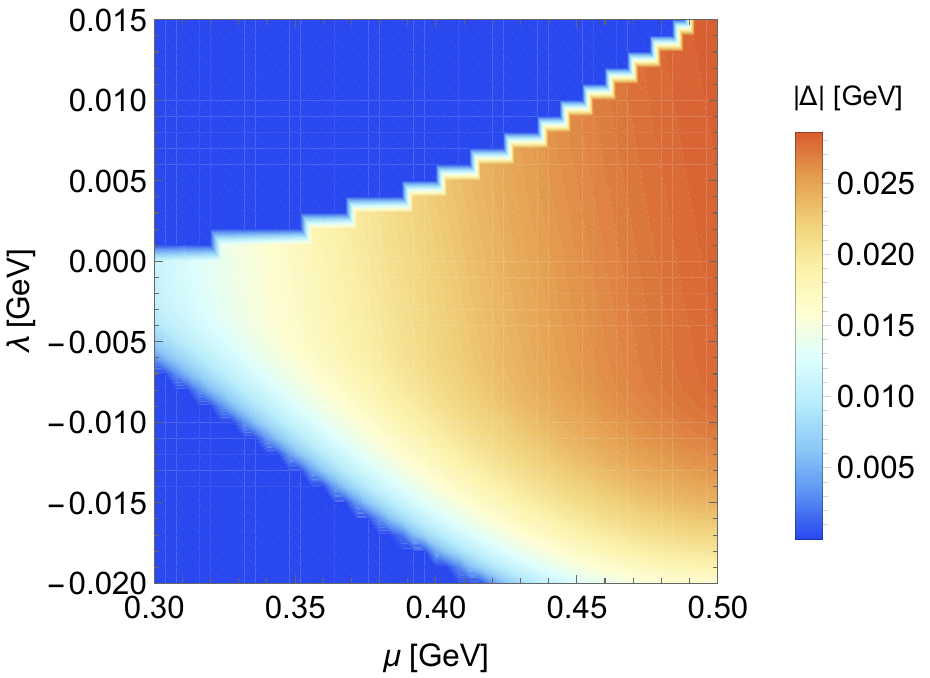}
\caption{The gap $|\Delta|$ as a function of $\mu$ and $\lambda$ at $T=0$ GeV.}
\label{fig:delta}\vspace{-5mm}
\end{center}
\end{figure}

\section{Results}
\label{sec:results}

Let us consider the case of zero temperature ($T\!=\!0$).
Before 
presenting numerical results, 
it is instructive to
investigate the approximate analytic solution for the gap at $\lambda = 0$.
From Eq.~(\ref{eq:thermo_pot}), we obtain the gap equation
\begin{eqnarray}
\Delta = \frac{1}{2}N_{c} G_{c} \int_{0}^{\Lambda} 
\frac{k^{2}\mathrm{d}k}{2\pi^{2}}
\frac{\Delta}{\sqrt{(k-\mu)^{2}+8N_{f}|\Delta|^{2}}} .
 \label{eq:gap_eq}
\end{eqnarray}
Notice that the gap equation (\ref{eq:gap_eq}) provided by the energy-momentum dispersion, $E_{-}(k)$ in Eq.~(\ref{eq:eigenmode1}), contains contribution only from the states below the Fermi surface, as seen in the middle panel ($\lambda=0$) of Fig.~\ref{fig:dispersion}.
We also notice that the states with the momentum around $k \simeq \mu$ contributes dominantly to the gap equation, and that the contributions from the states lying deeply inside the Fermi surface are relatively suppressed.
The gap equation (\ref{eq:gap_eq}) gives two solutions: $|\Delta|=0$ and
\begin{eqnarray}
 |\Delta| 
 &\simeq&
  \alpha \sqrt{\frac{\mu(\Lambda-\mu)}{2N_{f}}} \exp\left(-\frac{2\pi^2}{N_{c}\,\mu^{2}G_{c}}\right),
 \label{eq:gap_approx}
\end{eqnarray} 
where $\alpha$ is a factor independent of $G_c$ and is approximately given by $\alpha=\exp\left( {(\Lambda^{2}+2\Lambda\mu-6\mu^{2})}/{4\mu^{2}} \right)$ for a small $|\Delta|$.
The latter solution gives the most stable state.
Importantly, the finite gap always exists for any small coupling constant $G_{c}>0$.
This suggests that the gap appears even in the weak coupling regime, which is similar to the Cooper instability of the superconductivity.
It is also interesting to notice that the gap contains the exponential factor $\exp \left(-2\pi^2/(N_c\,\mu^2 G_c)\right)$ which is common with the factor appearing in the Kondo scale~\cite{Yasui:2013xr}, and thus increases with increasing coupling strength. 

The gap equation $\frac{\partial}{\partial \Delta^{\ast}} \Omega(T=0,\mu,\lambda;\Delta) = 0$ for finite values of $\mu$ and $\lambda$ must be solved numerically. We consider $N_{f}=2$ in the following. The result for the gap $|\Delta|$ as a function of $\mu$ and $\lambda$ is shown in Fig.~\ref{fig:delta}, which essentially corresponds to the phase diagram on the $\mu$-$\lambda$ plane. We find that $|\Delta|$ increases as $\mu$ increases. This is reasonable because the Fermi surface area, and thus the density of states at the Fermi surface, becomes larger with the increasing Fermi momentum. 
We have checked that the analytic solution (\ref{eq:gap_approx}) is relatively a good approximation to the numerical result at least for $\lambda= 0$ GeV. For example, Eq.~(\ref{eq:gap_approx}) gives $|\Delta|=0.026$ GeV at $\mu=0.5$ GeV, which is approximately consistent with the full numerical result, $|\Delta|=0.028$ GeV. 
The existence of a finite gap indicates that the ground state of the matter is not the normal phase but 
 the one with mixing between a light quark and a heavy quark, which we 
 call the ``Kondo phase".

\begin{figure}[t]
\begin{center}
  \begin{minipage}[b]{1.0\linewidth}
    \includegraphics[keepaspectratio, scale=0.45]{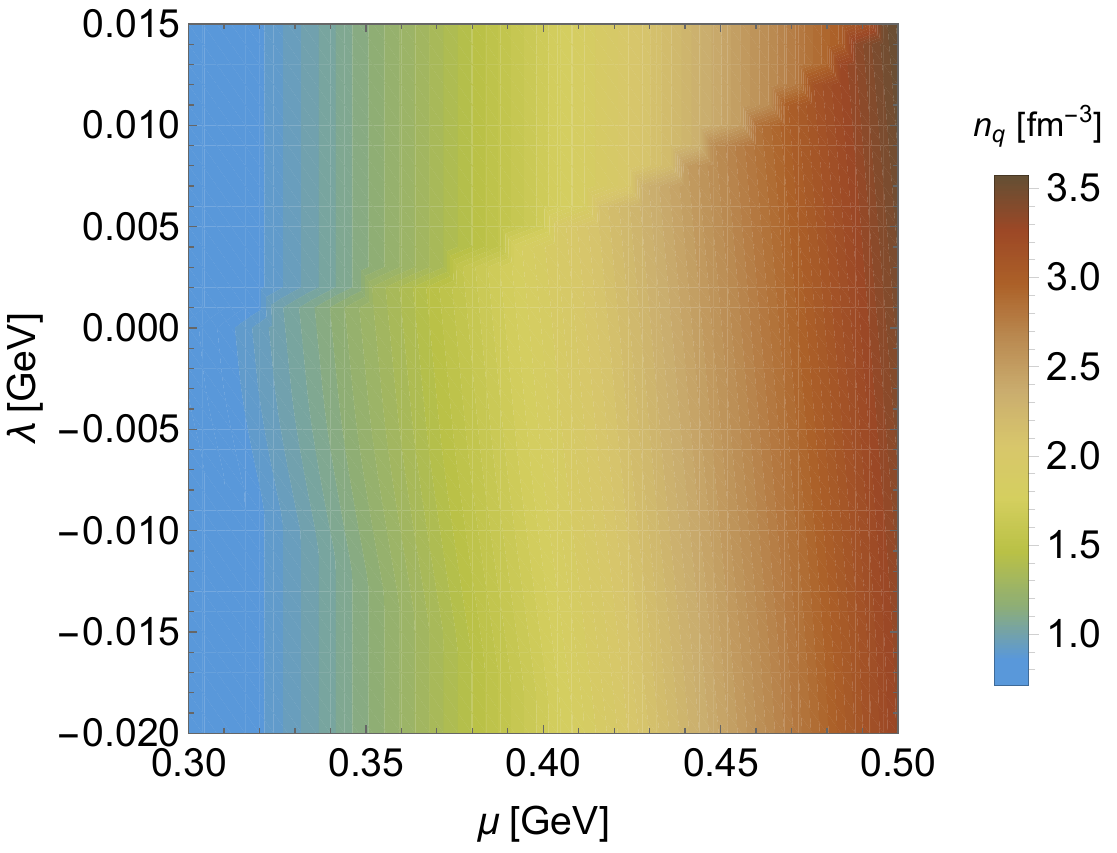}
  \end{minipage} \\
  \begin{minipage}[b]{1.0\linewidth}
    \includegraphics[keepaspectratio, scale=0.45]{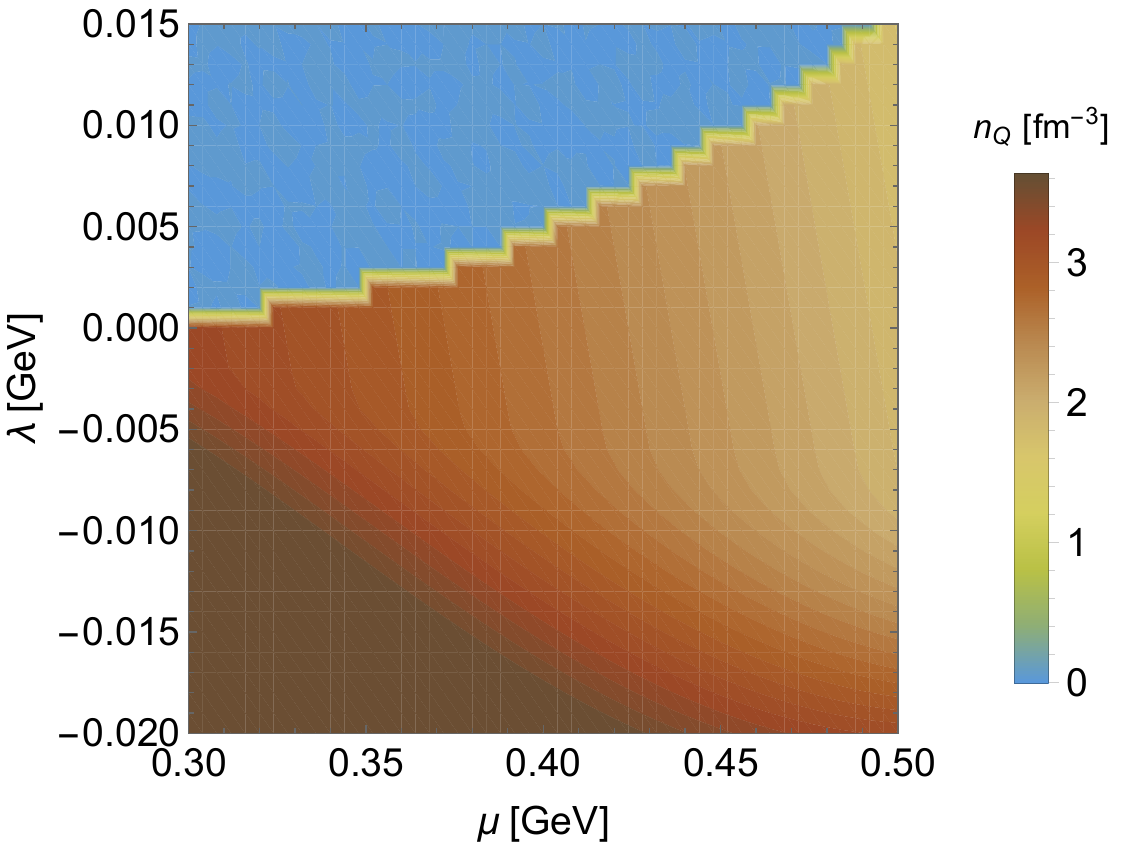}
  \end{minipage}
  \caption{The number density of light quark $n_{q}$ and heavy quark $n_{Q}$ as functions of $\mu$ and $\lambda$ at $T=0$ GeV.} 
  \label{fig:light_heavy}\vspace{-5mm}
  \end{center}
\end{figure}

In Fig.~\ref{fig:light_heavy}, the number densities of light quarks and heavy quarks, $n_{q}$ and $n_{Q}$, are shown. They are not control parameters but must be dynamically determined through the thermodynamical potential.
For $\mu=0.5$ GeV, we obtain $n_{q}=3.5$ fm$^{-3}$ and $n_{Q}=1.8$ fm$^{-3}$ at $\lambda=0.01$ GeV, and $n_{q}=3.4$ fm$^{-3}$ and $n_{Q}=2.0$ fm$^{-3}$ at $\lambda=-0.01$ GeV.

\begin{figure}[tbp]
\begin{center}
\includegraphics[scale=0.45]{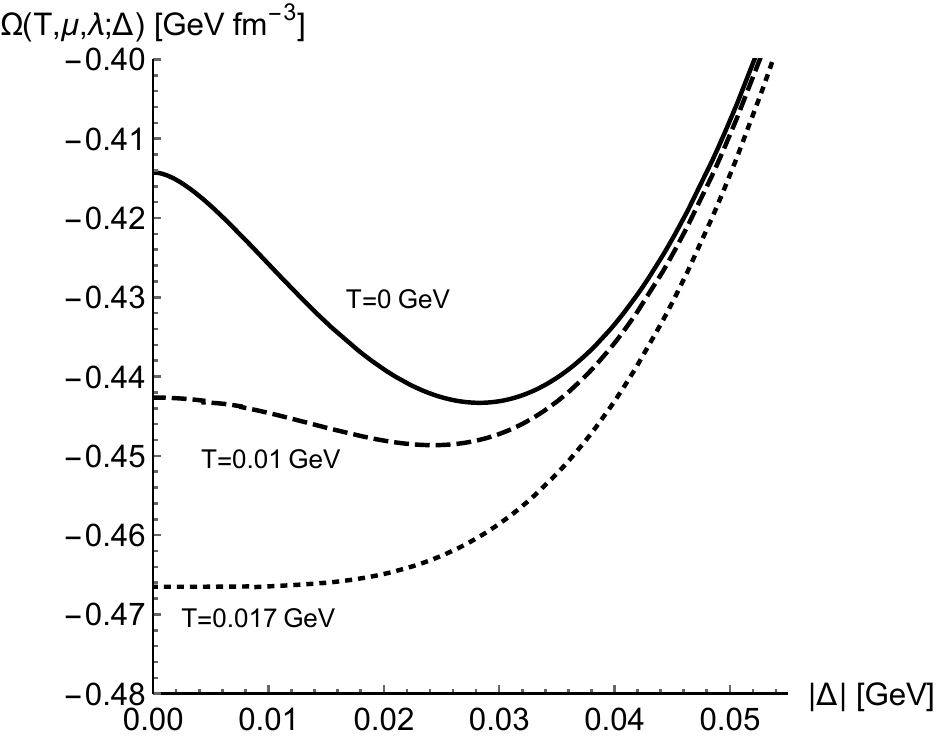}
\caption{The thermodynamic potential $\Omega(T,\mu,\lambda;\Delta)$ as a function of $|\Delta|$ with $\mu=0.5$ GeV and $\lambda=0$ GeV for several temperatures.}\vspace{-5mm}
\label{fig:thermo_pot_temp}
\end{center}
\end{figure}

The thermodynamic potentials $\Omega(T,\mu,\lambda;\Delta)$ at finite temperatures ($T \neq 0$) are plotted as functions of $|\Delta|$ in Fig.~\ref{fig:thermo_pot_temp}.
Fixing $\mu=0.5$ GeV and $\lambda = 0$ GeV, we have $|\Delta|=0.028$ GeV at $T=0$ GeV.
The gap decreases with increasing temperature, e.g. $|\Delta|=0.024$ GeV at $T=0.01$ GeV, and it becomes zero at 
$T=0.017$ GeV.

The system prefers to form a finite gap when the crossing point $(E,k)=(\lambda,\lambda+\mu)$ of the two original dispersions, $E=k-\mu$, $E=\lambda$, is close to the Fermi surface (see Fig.~\ref{fig:dispersion}). This is because a finite gap reduces the total energy most effectively when two new dispersions $E_\pm(k)$ are asymmetrically involved in the Fermi sea. This expectation is indeed confirmed in the numerical result. Figure~\ref{fig:delta} shows that when $\lambda$ becomes positively/negatively large, the gap $|\Delta|$ becomes small.

The results for $|\Delta|$, $n_q$, and $n_Q$ have a similar structure with a sudden change in the positive $\lambda$ region (though it is faint for $n_q$). Figures~\ref{fig:delta} and \ref{fig:light_heavy} show that the (uniform) gap is formed when the heavy quark density is high as long as $|\lambda|$ is close to zero (the Fermi surface). This is reasonable because an uniform condensate will be realized when the correlation/coherence length $\ell \sim 1/|\Delta|$ is much longer than the averaged inter-heavy-quark distance $d\sim 1/n_Q^{1/3}$. When $n_Q$ is small enough, it becomes difficult to maintain a spatially uniform gap. Therefore there is a close correlation between the behaviors of $|\Delta|$ and $n_Q$.

\begin{figure}[t]
\begin{center}
\includegraphics[scale=0.45]{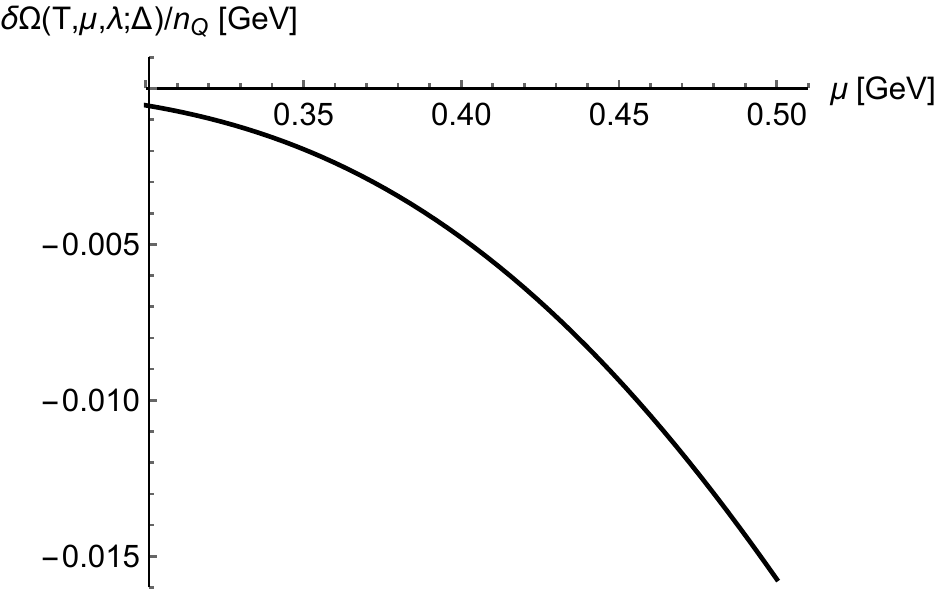}
\caption{The energy gain of quark matter per heavy quark as a function of $\mu$ at $\lambda=0$ GeV at $T=0$ GeV.}
\label{fig:energy_per_heavy}\vspace{-5mm}
\end{center}
\end{figure}

So far, we have assumed the uniformity of heavy quark distribution as an ideal situation.
As we mentioned in the Introduction, what we expect in reality is the presence of droplets with finite volumes in which heavy impurities are gathered at relatively high densities.
One can re-interpret the obtained results by defining the energy gain per a single heavy quark. Such quantity can be used to estimate the energy gain of  a droplet if its spatial size is much larger than the coherence length $\ell$.
In the present scheme, the energy gain per single heavy quark can be estimated as $\delta \Omega(T,\mu,\lambda;\Delta)/n_{Q}$ with $\delta \Omega(T,\mu,\lambda;\Delta)=\Omega(T,\mu,\lambda;\Delta)-\Omega(T,\mu,\lambda;0)$ as shown as a function of $\mu$ at $\lambda=0$ GeV and $T=0$ GeV in Fig.~\ref{fig:energy_per_heavy}.
As a numerical value, we obtain $\delta \Omega(T,\mu,\lambda;\Delta)/n_{Q}=-0.016$ GeV at $\mu=0.5$ GeV, $\lambda=0$ GeV and $T=0$ GeV.

Finally, we comment on the effect of several inter-quark correlations which are not taken into account in the present analysis. At high density, the attraction between two light quarks on the Fermi surface leads to the color superconductivity in quark matter. At relatively lower density, the attraction between a light quark and a light antiquark can form the chiral condensate resulting in the chiral symmetry breaking. Those correlations are competitive to the QCD Kondo effect.
For example, it was reported that the Kondo phase is excluded by a sufficiently large gap
in the two-flavor color-superconducting (2SC) phase~\cite{Kanazawa:2016ihl}.
As mentioned in Ref.~\cite{Hattori:2015hka}, however, for three-flavor QCD, since strange ($s$) quarks are decoupled from the formation of the Cooper pair in the 2SC phase, the color-exchange interaction between an $s$ quark and a heavy quark can lead to the QCD Kondo effect.
On the other hand, it was shown that the Kondo phase coexists with the chiral condensate~\cite{Suzuki:2017gde}.
As the other possible competition, one may think of the heavy-light ``diquark" channel. However, this is not color-singlet, and its contribution can be relatively suppressed as expected in the large $N_{c}$ limit.
In fact, it was shown for the single heavy quark case that the heavy-light diquark channel is unfavored not only in the large $N_{c}$ limit but also at $N_{c}=3$~\cite{Yasui:2016yet}.

In those competitions, the symmetry breaking pattern should be different from Eqs.~(\ref{eq:group_G}) and (\ref{eq:group_H}).
For example, let us consider the case where the Kondo phase coexists with the chiral condensate.
Supposing that the chiral condensate for all light flavors, we have the symmetry $G=\mathrm{SO}(3)_{\mathrm{space}} \times \mathrm{SU}(2)_{\mathrm{HQS}} \times \mathrm{U}(1)_{\mathrm{Q}} \times \mathrm{U}(1)_{\mathrm{V}} \times \mathrm{SU}(N_{f})_{\mathrm{R}+\mathrm{L}}$ in the ground state before switching on the Kondo phase. 
Here we did not include $\mathrm{U}(1)_{\mathrm{A}}$ which will be anomalously broken in QCD.
After the Kondo phase is switched on, the symmetry will be broken to $H=\mathrm{SO}(3)_{\mathrm{space}} \times \mathrm{U}(1)_{\mathrm{Q}+\mathrm{V}} 
\times \mathrm{SU}(N_{f}-1)_{\mathrm{R}+\mathrm{L}}$.

\section{Summary and outlook}
\label{sec:summary}
We discussed the QCD Kondo effect in the quark matter with heavy quarks as impurity particles.
We introduced the color exchange interaction and applied the mean-field approximation to the condensate composed of a light quark and a heavy quark as the Kondo cloud. 
We found the Kondo phase where the finite condensate is formed as the Kondo cloud.

In future, to go beyond the mean-field approximation, we can apply the random-phase  approximation (RPA) as studied in Ref.~\cite{Yasui:2016ngy}. This is important because the gap is associated with the formation of a bound state, which is described by the fluctuation around the ground state within the RPA.
It is also interesting to apply other non-perturbative approaches developed in condensed matter physics, such as the numerical renormalization group~\cite{Wilson:1974mb}, 
the energy-variation, the Green's function method and the Bethe ansatz (see Refs.~\cite{Hewson,Yosida,Yamada}). 

Furthermore, the present analysis does not include the correlations between light quarks, such as the quark-(anti)quark condensate for the color superconductivity and the spontaneous chiral symmetry breaking.
Such situations can be studied as an extension of the current approach by using the mean-field theory
and the many-body techniques beyond that.
In the present discussion, we considered only the homogeneous state for the number density of the heavy quarks. It will be worth to consider the possibility of realization of inhomogeneous state which may be similar to the chiral density wave and the LOFF state in superconductivity (see e.g. Ref.~\cite{Kojo:2009ha}).

The QCD Kondo effect may be analogous to the Kondo effect for Dirac fermions in condensed matter of electrons~\cite{AONO:2013,Yanagisawa:2015conf,Yanagisawa:2015}~\footnote{See, for example, Refs.~\cite{Novoselov:2005,Zhang:2005,Hsieh:2008,Mitchell:2013,KANAO:2012} for recent studies of Dirac fermions.}.
The Kondo effect in Majorana fermions in topological matter is also discussed~\cite{Beri:2012,Altland:2014,Cheng:2014,Eriksson:2014}.
In addition, it was recently shown that the QCD Kondo effect emerges for the degenerate states in the Landau level under magnetic field instead of the Fermi surface~\cite{Ozaki:2015sya}.
The Kondo phase diagram in various environments might be interesting.
Those subjects will be left for future works.

\section*{Acknowledgments}
The authors would like to sincerely thank T.~Kanazawa for pointing out the correction in the energy-momentum dispersion relations.
This work is supported by the Grant-in-Aid for Scientific Research (Grant No.~25247036, No.~15K17641 and No.~16K05366) from Japan Society for the Promotion of Science (JSPS), and is also supported partly by the Center for the Promotion of Integrated Sciences (CPIS) of Sokendai.
The authors thank the Yukawa Institute for Theoretical Physics, Kyoto University, 
where this work was partially carried out during the YITP-T-15-08 on 
``Exotic hadrons from high energy collisions". 

\bibliography{reference}

\begin{thebibliography}{52}%
\makeatletter
\providecommand \@ifxundefined [1]{%
 \@ifx{#1\undefined}
}%
\providecommand \@ifnum [1]{%
 \ifnum #1\expandafter \@firstoftwo
 \else \expandafter \@secondoftwo
 \fi
}%
\providecommand \@ifx [1]{%
 \ifx #1\expandafter \@firstoftwo
 \else \expandafter \@secondoftwo
 \fi
}%
\providecommand \natexlab [1]{#1}%
\providecommand \enquote  [1]{``#1''}%
\providecommand \bibnamefont  [1]{#1}%
\providecommand \bibfnamefont [1]{#1}%
\providecommand \citenamefont [1]{#1}%
\providecommand \href@noop [0]{\@secondoftwo}%
\providecommand \href [0]{\begingroup \@sanitize@url \@href}%
\providecommand \@href[1]{\@@startlink{#1}\@@href}%
\providecommand \@@href[1]{\endgroup#1\@@endlink}%
\providecommand \@sanitize@url [0]{\catcode `\\12\catcode `\$12\catcode
  `\&12\catcode `\#12\catcode `\^12\catcode `\_12\catcode `\%12\relax}%
\providecommand \@@startlink[1]{}%
\providecommand \@@endlink[0]{}%
\providecommand \url  [0]{\begingroup\@sanitize@url \@url }%
\providecommand \@url [1]{\endgroup\@href {#1}{\urlprefix }}%
\providecommand \urlprefix  [0]{URL }%
\providecommand \Eprint [0]{\href }%
\providecommand \doibase [0]{http://dx.doi.org/}%
\providecommand \selectlanguage [0]{\@gobble}%
\providecommand \bibinfo  [0]{\@secondoftwo}%
\providecommand \bibfield  [0]{\@secondoftwo}%
\providecommand \translation [1]{[#1]}%
\providecommand \BibitemOpen [0]{}%
\providecommand \bibitemStop [0]{}%
\providecommand \bibitemNoStop [0]{.\EOS\space}%
\providecommand \EOS [0]{\spacefactor3000\relax}%
\providecommand \BibitemShut  [1]{\csname bibitem#1\endcsname}%
\let\auto@bib@innerbib\@empty
%</preamble>
\bibitem [{\citenamefont {Kondo}(1964)}]{Kondo:1964}%
  \BibitemOpen
  \bibfield  {author} {\bibinfo {author} {\bibfnamefont {J.}~\bibnamefont
  {Kondo}},\ }\href {\doibase 10.1143/PTP.32.37} {\bibfield  {journal}
  {\bibinfo  {journal} {Prog. Theor. Phys.}\ }\textbf {\bibinfo {volume}
  {32}},\ \bibinfo {pages} {37} (\bibinfo {year} {1964})}\BibitemShut {NoStop}%
%%CITATION = PTPKA,32,37;%%
\bibitem [{\citenamefont {Hewson}(1993)}]{Hewson}%
  \BibitemOpen
  \bibfield  {author} {\bibinfo {author} {\bibfnamefont {A.~C.}\ \bibnamefont
  {Hewson}},\ }\href@noop {} {\emph {\bibinfo {title} {{The Kondo Problem to
  Heavy Fermionss}}}}\ (\bibinfo  {publisher} {Cambridge University Press},\
  \bibinfo {year} {1993})\BibitemShut {NoStop}%
\bibitem [{\citenamefont {Yosida}(1996)}]{Yosida}%
  \BibitemOpen
  \bibfield  {author} {\bibinfo {author} {\bibfnamefont {K.}~\bibnamefont
  {Yosida}},\ }\href@noop {} {\emph {\bibinfo {title} {{Theory of
  Magnetism}}}}\ (\bibinfo  {publisher} {Springer-Verlag Berlin Heidelberg},\
  \bibinfo {year} {1996})\BibitemShut {NoStop}%
\bibitem [{\citenamefont {Yamada}(2004)}]{Yamada}%
  \BibitemOpen
  \bibfield  {author} {\bibinfo {author} {\bibfnamefont {K.}~\bibnamefont
  {Yamada}},\ }\href@noop {} {\emph {\bibinfo {title} {{Electron Correlation in
  Metals}}}}\ (\bibinfo  {publisher} {Cambridge University Press},\ \bibinfo
  {year} {2004})\BibitemShut {NoStop}%
\bibitem [{\citenamefont {Goldhaber-Gordon}\ \emph {et~al.}(1998)\citenamefont
  {Goldhaber-Gordon}, \citenamefont {Shtrikman}, \citenamefont {Mahalu},
  \citenamefont {Abusch-Magder}, \citenamefont {Meirav},\ and\ \citenamefont
  {Kastner}}]{Goldhaber-Gordon:1998}%
  \BibitemOpen
  \bibfield  {author} {\bibinfo {author} {\bibfnamefont {D.}~\bibnamefont
  {Goldhaber-Gordon}}, \bibinfo {author} {\bibfnamefont {H.}~\bibnamefont
  {Shtrikman}}, \bibinfo {author} {\bibfnamefont {D.}~\bibnamefont {Mahalu}},
  \bibinfo {author} {\bibfnamefont {D.}~\bibnamefont {Abusch-Magder}}, \bibinfo
  {author} {\bibfnamefont {U.}~\bibnamefont {Meirav}}, \ and\ \bibinfo {author}
  {\bibfnamefont {M.~A.}\ \bibnamefont {Kastner}},\ }\href {\doibase
  10.1038/34373} {\bibfield  {journal} {\bibinfo  {journal} {Nature}\ }\textbf
  {\bibinfo {volume} {391}},\ \bibinfo {pages} {156} (\bibinfo {year}
  {1998})}\BibitemShut {NoStop}%
\bibitem [{\citenamefont {Cronenwett}\ \emph {et~al.}(1998)\citenamefont
  {Cronenwett}, \citenamefont {Oosterkamp},\ and\ \citenamefont
  {Kouwenhoven}}]{Cronenwett:1998}%
  \BibitemOpen
  \bibfield  {author} {\bibinfo {author} {\bibfnamefont {S.~M.}\ \bibnamefont
  {Cronenwett}}, \bibinfo {author} {\bibfnamefont {T.~H.}\ \bibnamefont
  {Oosterkamp}}, \ and\ \bibinfo {author} {\bibfnamefont {L.~P.}\ \bibnamefont
  {Kouwenhoven}},\ }\href {\doibase 10.1126/science.281.5376.540} {\bibfield
  {journal} {\bibinfo  {journal} {Science}\ }\textbf {\bibinfo {volume}
  {281}},\ \bibinfo {pages} {540} (\bibinfo {year} {1998})},\ \Eprint
  {http://arxiv.org/abs/9804211} {arXiv:9804211 [cond-mat]} \BibitemShut
  {NoStop}%
\bibitem [{\citenamefont {van~der Wiel}\ \emph {et~al.}(2000)\citenamefont
  {van~der Wiel}, \citenamefont {Franceschi}, \citenamefont {Fujisawa},
  \citenamefont {Elzerman}, \citenamefont {Tarucha},\ and\ \citenamefont
  {Kouwenhoven}}]{Wiel:2000}%
  \BibitemOpen
  \bibfield  {author} {\bibinfo {author} {\bibfnamefont {W.~G.}\ \bibnamefont
  {van~der Wiel}}, \bibinfo {author} {\bibfnamefont {S.~D.}\ \bibnamefont
  {Franceschi}}, \bibinfo {author} {\bibfnamefont {T.}~\bibnamefont
  {Fujisawa}}, \bibinfo {author} {\bibfnamefont {J.~M.}\ \bibnamefont
  {Elzerman}}, \bibinfo {author} {\bibfnamefont {S.}~\bibnamefont {Tarucha}}, \
  and\ \bibinfo {author} {\bibfnamefont {L.~P.}\ \bibnamefont {Kouwenhoven}},\
  }\href {\doibase 10.1126/science.289.5487.2105} {\bibfield  {journal}
  {\bibinfo  {journal} {Science}\ }\textbf {\bibinfo {volume} {289}},\ \bibinfo
  {pages} {2105} (\bibinfo {year} {2000})},\ \Eprint
  {http://arxiv.org/abs/0009405} {arXiv:0009405 [cond-mat]} \BibitemShut
  {NoStop}%
\bibitem [{\citenamefont {Jeong}\ \emph {et~al.}(2001)\citenamefont {Jeong},
  \citenamefont {Chang},\ and\ \citenamefont {Melloch}}]{Jeong:2001}%
  \BibitemOpen
  \bibfield  {author} {\bibinfo {author} {\bibfnamefont {H.}~\bibnamefont
  {Jeong}}, \bibinfo {author} {\bibfnamefont {A.~M.}\ \bibnamefont {Chang}}, \
  and\ \bibinfo {author} {\bibfnamefont {M.~R.}\ \bibnamefont {Melloch}},\
  }\href {\doibase 10.1126/science.1063182} {\bibfield  {journal} {\bibinfo
  {journal} {Science}\ }\textbf {\bibinfo {volume} {293}},\ \bibinfo {pages}
  {2221} (\bibinfo {year} {2001})}\BibitemShut {NoStop}%
\bibitem [{\citenamefont {Park}\ \emph {et~al.}(2002)\citenamefont {Park},
  \citenamefont {Pasupathy}, \citenamefont {Goldsmith}, \citenamefont {Chang},
  \citenamefont {Yaish}, \citenamefont {Petta}, \citenamefont {Rinkoski},
  \citenamefont {Sethna}, \citenamefont {Abru{\~n}a}, \citenamefont {McEuen},\
  and\ \citenamefont {Ralph}}]{Park:2002}%
  \BibitemOpen
  \bibfield  {author} {\bibinfo {author} {\bibfnamefont {J.}~\bibnamefont
  {Park}}, \bibinfo {author} {\bibfnamefont {A.~N.}\ \bibnamefont {Pasupathy}},
  \bibinfo {author} {\bibfnamefont {J.~I.}\ \bibnamefont {Goldsmith}}, \bibinfo
  {author} {\bibfnamefont {C.}~\bibnamefont {Chang}}, \bibinfo {author}
  {\bibfnamefont {Y.}~\bibnamefont {Yaish}}, \bibinfo {author} {\bibfnamefont
  {J.~R.}\ \bibnamefont {Petta}}, \bibinfo {author} {\bibfnamefont
  {M.}~\bibnamefont {Rinkoski}}, \bibinfo {author} {\bibfnamefont {J.~P.}\
  \bibnamefont {Sethna}}, \bibinfo {author} {\bibfnamefont {H.~D.}\
  \bibnamefont {Abru{\~n}a}}, \bibinfo {author} {\bibfnamefont {P.~L.}\
  \bibnamefont {McEuen}}, \ and\ \bibinfo {author} {\bibfnamefont {D.~C.}\
  \bibnamefont {Ralph}},\ }\href {\doibase 10.1038/nature00791} {\bibfield
  {journal} {\bibinfo  {journal} {Nature}\ }\textbf {\bibinfo {volume} {417}},\
  \bibinfo {pages} {722} (\bibinfo {year} {2002})}\BibitemShut {NoStop}%
\bibitem [{\citenamefont {Hur}(2015)}]{Hur:2015}%
  \BibitemOpen
  \bibfield  {author} {\bibinfo {author} {\bibfnamefont {K.~L.}\ \bibnamefont
  {Hur}},\ }\href {\doibase 10.1038/526203a} {\bibfield  {journal} {\bibinfo
  {journal} {Nature}\ }\textbf {\bibinfo {volume} {526}},\ \bibinfo {pages}
  {203} (\bibinfo {year} {2015})}\BibitemShut {NoStop}%
\bibitem [{\citenamefont {Balatsky}\ \emph {et~al.}(2006)\citenamefont
  {Balatsky}, \citenamefont {Vekhter},\ and\ \citenamefont
  {Zhu}}]{Balatsky:2006}%
  \BibitemOpen
  \bibfield  {author} {\bibinfo {author} {\bibfnamefont {A.~V.}\ \bibnamefont
  {Balatsky}}, \bibinfo {author} {\bibfnamefont {I.}~\bibnamefont {Vekhter}}, \
  and\ \bibinfo {author} {\bibfnamefont {J.-X.}\ \bibnamefont {Zhu}},\ }\href
  {\doibase 10.1103/RevModPhys.78.373} {\bibfield  {journal} {\bibinfo
  {journal} {Rev. Mod. Phys. B}\ }\textbf {\bibinfo {volume} {78}},\ \bibinfo
  {pages} {373} (\bibinfo {year} {2006})},\ \Eprint
  {http://arxiv.org/abs/cond-mat/0411318} {arXiv:cond-mat/0411318
  [cond-mat.supr-con]} \BibitemShut {NoStop}%
\bibitem [{\citenamefont {Nishida}(2013)}]{PhysRevLett.111.135301}%
  \BibitemOpen
  \bibfield  {author} {\bibinfo {author} {\bibfnamefont {Y.}~\bibnamefont
  {Nishida}},\ }\href {\doibase 10.1103/PhysRevLett.111.135301} {\bibfield
  {journal} {\bibinfo  {journal} {Phys. Rev. Lett.}\ }\textbf {\bibinfo
  {volume} {111}},\ \bibinfo {pages} {135301} (\bibinfo {year}
  {2013})}\BibitemShut {NoStop}%
\bibitem [{\citenamefont {Yasui}\ and\ \citenamefont
  {Sudoh}(2013)}]{Yasui:2013xr}%
  \BibitemOpen
  \bibfield  {author} {\bibinfo {author} {\bibfnamefont {S.}~\bibnamefont
  {Yasui}}\ and\ \bibinfo {author} {\bibfnamefont {K.}~\bibnamefont {Sudoh}},\
  }\href {\doibase 10.1103/PhysRevC.88.015201} {\bibfield  {journal} {\bibinfo
  {journal} {Phys. Rev. C}\ }\textbf {\bibinfo {volume} {88}},\ \bibinfo
  {pages} {015201} (\bibinfo {year} {2013})},\ \Eprint
  {http://arxiv.org/abs/1301.6830} {arXiv:1301.6830 [hep-ph]} \BibitemShut
  {NoStop}%
%%CITATION = ARXIV:1301.6830;%%
\bibitem [{\citenamefont {Hattori}\ \emph {et~al.}(2015)\citenamefont
  {Hattori}, \citenamefont {Itakura}, \citenamefont {Ozaki},\ and\
  \citenamefont {Yasui}}]{Hattori:2015hka}%
  \BibitemOpen
  \bibfield  {author} {\bibinfo {author} {\bibfnamefont {K.}~\bibnamefont
  {Hattori}}, \bibinfo {author} {\bibfnamefont {K.}~\bibnamefont {Itakura}},
  \bibinfo {author} {\bibfnamefont {S.}~\bibnamefont {Ozaki}}, \ and\ \bibinfo
  {author} {\bibfnamefont {S.}~\bibnamefont {Yasui}},\ }\href {\doibase
  10.1103/PhysRevD.92.065003} {\bibfield  {journal} {\bibinfo  {journal} {Phys.
  Rev. D}\ }\textbf {\bibinfo {volume} {92}},\ \bibinfo {pages} {065003}
  (\bibinfo {year} {2015})},\ \Eprint {http://arxiv.org/abs/1504.07619}
  {arXiv:1504.07619 [hep-ph]} \BibitemShut {NoStop}%
%%CITATION = ARXIV:1504.07619;%%
\bibitem [{\citenamefont {Ozaki}\ \emph {et~al.}(2016)\citenamefont {Ozaki},
  \citenamefont {Itakura},\ and\ \citenamefont {Kuramoto}}]{Ozaki:2015sya}%
  \BibitemOpen
  \bibfield  {author} {\bibinfo {author} {\bibfnamefont {S.}~\bibnamefont
  {Ozaki}}, \bibinfo {author} {\bibfnamefont {K.}~\bibnamefont {Itakura}}, \
  and\ \bibinfo {author} {\bibfnamefont {Y.}~\bibnamefont {Kuramoto}},\ }\href
  {\doibase 10.1103/PhysRevD.94.074013} {\bibfield  {journal} {\bibinfo
  {journal} {Phys. Rev.}\ }\textbf {\bibinfo {volume} {D94}},\ \bibinfo {pages}
  {074013} (\bibinfo {year} {2016})},\ \Eprint
  {http://arxiv.org/abs/1509.06966} {arXiv:1509.06966 [hep-ph]} \BibitemShut
  {NoStop}%
%%CITATION = ARXIV:1509.06966;%%
\bibitem [{\citenamefont {Nambu}\ and\ \citenamefont
  {Jona-Lasinio}(1961)}]{Nambu:1961tp}%
  \BibitemOpen
  \bibfield  {author} {\bibinfo {author} {\bibfnamefont {Y.}~\bibnamefont
  {Nambu}}\ and\ \bibinfo {author} {\bibfnamefont {G.}~\bibnamefont
  {Jona-Lasinio}},\ }\href {\doibase 10.1103/PhysRev.122.345} {\bibfield
  {journal} {\bibinfo  {journal} {Phys. Rev.}\ }\textbf {\bibinfo {volume}
  {122}},\ \bibinfo {pages} {345} (\bibinfo {year} {1961})}\BibitemShut
  {NoStop}%
%%CITATION = PHRVA,122,345;%%
\bibitem [{\citenamefont {Fukushima}\ and\ \citenamefont
  {Hatsuda}(2011)}]{Fukushima:2010bq}%
  \BibitemOpen
  \bibfield  {author} {\bibinfo {author} {\bibfnamefont {K.}~\bibnamefont
  {Fukushima}}\ and\ \bibinfo {author} {\bibfnamefont {T.}~\bibnamefont
  {Hatsuda}},\ }\href {\doibase 10.1088/0034-4885/74/1/014001} {\bibfield
  {journal} {\bibinfo  {journal} {Rept. Prog. Phys.}\ }\textbf {\bibinfo
  {volume} {74}},\ \bibinfo {pages} {014001} (\bibinfo {year} {2011})},\
  \Eprint {http://arxiv.org/abs/1005.4814} {arXiv:1005.4814 [hep-ph]}
  \BibitemShut {NoStop}%
%%CITATION = ARXIV:1005.4814;%%
\bibitem [{\citenamefont {Fukushima}\ and\ \citenamefont
  {Sasaki}(2013)}]{Fukushima:2013rx}%
  \BibitemOpen
  \bibfield  {author} {\bibinfo {author} {\bibfnamefont {K.}~\bibnamefont
  {Fukushima}}\ and\ \bibinfo {author} {\bibfnamefont {C.}~\bibnamefont
  {Sasaki}},\ }\href {\doibase 10.1016/j.ppnp.2013.05.003} {\bibfield
  {journal} {\bibinfo  {journal} {Prog. Part. Nucl. Phys.}\ }\textbf {\bibinfo
  {volume} {72}},\ \bibinfo {pages} {99} (\bibinfo {year} {2013})},\ \Eprint
  {http://arxiv.org/abs/1301.6377} {arXiv:1301.6377 [hep-ph]} \BibitemShut
  {NoStop}%
%%CITATION = ARXIV:1301.6377;%%
\bibitem [{\citenamefont {Alford}\ \emph {et~al.}(2008)\citenamefont {Alford},
  \citenamefont {Schmitt}, \citenamefont {Rajagopal},\ and\ \citenamefont
  {Sch{\"a}fer}}]{Alford:2007xm}%
  \BibitemOpen
  \bibfield  {author} {\bibinfo {author} {\bibfnamefont {M.~G.}\ \bibnamefont
  {Alford}}, \bibinfo {author} {\bibfnamefont {A.}~\bibnamefont {Schmitt}},
  \bibinfo {author} {\bibfnamefont {K.}~\bibnamefont {Rajagopal}}, \ and\
  \bibinfo {author} {\bibfnamefont {T.}~\bibnamefont {Sch{\"a}fer}},\ }\href
  {\doibase 10.1103/RevModPhys.80.1455} {\bibfield  {journal} {\bibinfo
  {journal} {Rev. Mod. Phys.}\ }\textbf {\bibinfo {volume} {80}},\ \bibinfo
  {pages} {1455} (\bibinfo {year} {2008})},\ \Eprint
  {http://arxiv.org/abs/0709.4635} {arXiv:0709.4635 [hep-ph]} \BibitemShut
  {NoStop}%
%%CITATION = ARXIV:0709.4635;%%
\bibitem [{\citenamefont {Takano}\ and\ \citenamefont
  {Ogawa}(1966)}]{Takano:1966}%
  \BibitemOpen
  \bibfield  {author} {\bibinfo {author} {\bibfnamefont {F.}~\bibnamefont
  {Takano}}\ and\ \bibinfo {author} {\bibfnamefont {T.}~\bibnamefont {Ogawa}},\
  }\href {\doibase 10.1143/PTP.35.343} {\bibfield  {journal} {\bibinfo
  {journal} {Prog. Theor. Phys.}\ }\textbf {\bibinfo {volume} {35}},\ \bibinfo
  {pages} {343} (\bibinfo {year} {1966})}\BibitemShut {NoStop}%
\bibitem [{\citenamefont {Yoshimori}\ and\ \citenamefont
  {Sakurai}(1970)}]{Yoshimori:1970}%
  \BibitemOpen
  \bibfield  {author} {\bibinfo {author} {\bibfnamefont {A.}~\bibnamefont
  {Yoshimori}}\ and\ \bibinfo {author} {\bibfnamefont {A.}~\bibnamefont
  {Sakurai}},\ }\href {\doibase 10.1143/PTPS.46.162} {\bibfield  {journal}
  {\bibinfo  {journal} {Prog. Theor. Phys. Suppl.}\ }\textbf {\bibinfo {volume}
  {46}},\ \bibinfo {pages} {162} (\bibinfo {year} {1970})}\BibitemShut
  {NoStop}%
\bibitem [{\citenamefont {Lacroix}\ and\ \citenamefont
  {Cyrot}(1979)}]{Lacroix:1979}%
  \BibitemOpen
  \bibfield  {author} {\bibinfo {author} {\bibfnamefont {C.}~\bibnamefont
  {Lacroix}}\ and\ \bibinfo {author} {\bibfnamefont {M.}~\bibnamefont
  {Cyrot}},\ }\href {\doibase 10.1103/PhysRevB.20.1969} {\bibfield  {journal}
  {\bibinfo  {journal} {Phys. Rev. B}\ }\textbf {\bibinfo {volume} {20}},\
  \bibinfo {pages} {1969} (\bibinfo {year} {1979})}\BibitemShut {NoStop}%
\bibitem [{\citenamefont {Coleman}(1983)}]{PhysRevB.28.5255}%
  \BibitemOpen
  \bibfield  {author} {\bibinfo {author} {\bibfnamefont {P.}~\bibnamefont
  {Coleman}},\ }\href {\doibase 10.1103/PhysRevB.28.5255} {\bibfield  {journal}
  {\bibinfo  {journal} {Phys. Rev. B}\ }\textbf {\bibinfo {volume} {28}},\
  \bibinfo {pages} {5255} (\bibinfo {year} {1983})}\BibitemShut {NoStop}%
\bibitem [{\citenamefont {Read}\ \emph {et~al.}(1984)\citenamefont {Read},
  \citenamefont {Newns},\ and\ \citenamefont {Doniach}}]{PhysRevB.30.3841}%
  \BibitemOpen
  \bibfield  {author} {\bibinfo {author} {\bibfnamefont {N.}~\bibnamefont
  {Read}}, \bibinfo {author} {\bibfnamefont {D.~M.}\ \bibnamefont {Newns}}, \
  and\ \bibinfo {author} {\bibfnamefont {S.}~\bibnamefont {Doniach}},\ }\href
  {\doibase 10.1103/PhysRevB.30.3841} {\bibfield  {journal} {\bibinfo
  {journal} {Phys. Rev. B}\ }\textbf {\bibinfo {volume} {30}},\ \bibinfo
  {pages} {3841} (\bibinfo {year} {1984})}\BibitemShut {NoStop}%
\bibitem [{\citenamefont {Eto}\ and\ \citenamefont {Nazarov}(2001)}]{Eto:2001}%
  \BibitemOpen
  \bibfield  {author} {\bibinfo {author} {\bibfnamefont {M.}~\bibnamefont
  {Eto}}\ and\ \bibinfo {author} {\bibfnamefont {Y.~V.}\ \bibnamefont
  {Nazarov}},\ }\href {\doibase 10.1103/PhysRevB.64.085322} {\bibfield
  {journal} {\bibinfo  {journal} {Phys. Rev. B}\ }\textbf {\bibinfo {volume}
  {64}},\ \bibinfo {pages} {085322} (\bibinfo {year} {2001})}\BibitemShut
  {NoStop}%
\bibitem [{\citenamefont
  {Yanagisawa}(2015{\natexlab{a}})}]{Yanagisawa:2015conf}%
  \BibitemOpen
  \bibfield  {author} {\bibinfo {author} {\bibfnamefont {T.}~\bibnamefont
  {Yanagisawa}},\ }\href {\doibase 10.1088/1742-6596/603/1/012014} {\bibfield
  {journal} {\bibinfo  {journal} {J. Phys.: Conference Series}\ }\textbf
  {\bibinfo {volume} {603}},\ \bibinfo {pages} {012014} (\bibinfo {year}
  {2015}{\natexlab{a}})},\ \Eprint {http://arxiv.org/abs/1502.07898}
  {arXiv:1502.07898 [cond-mat.str-el]} \BibitemShut {NoStop}%
\bibitem [{\citenamefont {Yanagisawa}(2015{\natexlab{b}})}]{Yanagisawa:2015}%
  \BibitemOpen
  \bibfield  {author} {\bibinfo {author} {\bibfnamefont {T.}~\bibnamefont
  {Yanagisawa}},\ }\href {\doibase 10.7566/JPSJ.84.074705} {\bibfield
  {journal} {\bibinfo  {journal} {J. Phys. Soc. Jpn.}\ }\textbf {\bibinfo
  {volume} {84}},\ \bibinfo {pages} {074705} (\bibinfo {year}
  {2015}{\natexlab{b}})},\ \Eprint {http://arxiv.org/abs/1505.05295}
  {arXiv:1505.05295 [cond-mat.str-el]} \BibitemShut {NoStop}%
\bibitem [{\citenamefont {Yasui}(2016)}]{Yasui:2016ngy}%
  \BibitemOpen
  \bibfield  {author} {\bibinfo {author} {\bibfnamefont {S.}~\bibnamefont
  {Yasui}},\ }\href {\doibase 10.1103/PhysRevC.93.065204} {\bibfield  {journal}
  {\bibinfo  {journal} {Phys. Rev.}\ }\textbf {\bibinfo {volume} {C93}},\
  \bibinfo {pages} {065204} (\bibinfo {year} {2016})},\ \Eprint
  {http://arxiv.org/abs/1602.00227} {arXiv:1602.00227 [hep-ph]} \BibitemShut
  {NoStop}%
%%CITATION = ARXIV:1602.00227;%%
\bibitem [{\citenamefont {Sugawara-Tanabe}\ and\ \citenamefont
  {Tanabe}(1979)}]{Sugawara-Tanabe:1979}%
  \BibitemOpen
  \bibfield  {author} {\bibinfo {author} {\bibfnamefont {K.}~\bibnamefont
  {Sugawara-Tanabe}}\ and\ \bibinfo {author} {\bibfnamefont {K.}~\bibnamefont
  {Tanabe}},\ }\href {\doibase 10.1103/PhysRevC.19.545} {\bibfield  {journal}
  {\bibinfo  {journal} {Phys. Rev. C}\ }\textbf {\bibinfo {volume} {19}},\
  \bibinfo {pages} {545} (\bibinfo {year} {1979})}\BibitemShut {NoStop}%
%%CITATION = PHRVA,C19,545;%%
\bibitem [{\citenamefont {Klimt}\ \emph {et~al.}(1990)\citenamefont {Klimt},
  \citenamefont {Lutz}, \citenamefont {Vogl},\ and\ \citenamefont
  {Weise}}]{Klimt:1989pm}%
  \BibitemOpen
  \bibfield  {author} {\bibinfo {author} {\bibfnamefont {S.}~\bibnamefont
  {Klimt}}, \bibinfo {author} {\bibfnamefont {M.~F.~M.}\ \bibnamefont {Lutz}},
  \bibinfo {author} {\bibfnamefont {U.}~\bibnamefont {Vogl}}, \ and\ \bibinfo
  {author} {\bibfnamefont {W.}~\bibnamefont {Weise}},\ }\href {\doibase
  10.1016/0375-9474(90)90123-4} {\bibfield  {journal} {\bibinfo  {journal}
  {Nucl. Phys.}\ }\textbf {\bibinfo {volume} {A516}},\ \bibinfo {pages} {429}
  (\bibinfo {year} {1990})}\BibitemShut {NoStop}%
%%CITATION = NUPHA,A516,429;%%
\bibitem [{\citenamefont {Vogl}\ \emph {et~al.}(1990)\citenamefont {Vogl},
  \citenamefont {Lutz}, \citenamefont {Klimt},\ and\ \citenamefont
  {Weise}}]{Vogl:1989ea}%
  \BibitemOpen
  \bibfield  {author} {\bibinfo {author} {\bibfnamefont {U.}~\bibnamefont
  {Vogl}}, \bibinfo {author} {\bibfnamefont {M.~F.~M.}\ \bibnamefont {Lutz}},
  \bibinfo {author} {\bibfnamefont {S.}~\bibnamefont {Klimt}}, \ and\ \bibinfo
  {author} {\bibfnamefont {W.}~\bibnamefont {Weise}},\ }\href {\doibase
  10.1016/0375-9474(90)90124-5} {\bibfield  {journal} {\bibinfo  {journal}
  {Nucl. Phys.}\ }\textbf {\bibinfo {volume} {A516}},\ \bibinfo {pages} {469}
  (\bibinfo {year} {1990})}\BibitemShut {NoStop}%
%%CITATION = NUPHA,A516,469;%%
\bibitem [{\citenamefont {Klevansky}(1992)}]{Klevansky:1992qe}%
  \BibitemOpen
  \bibfield  {author} {\bibinfo {author} {\bibfnamefont {S.~P.}\ \bibnamefont
  {Klevansky}},\ }\href {\doibase 10.1103/RevModPhys.64.649} {\bibfield
  {journal} {\bibinfo  {journal} {Rev. Mod. Phys.}\ }\textbf {\bibinfo {volume}
  {64}},\ \bibinfo {pages} {649} (\bibinfo {year} {1992})}\BibitemShut
  {NoStop}%
%%CITATION = RMPHA,64,649;%%
\bibitem [{\citenamefont {Hatsuda}\ and\ \citenamefont
  {Kunihiro}(1994)}]{Hatsuda:1994pi}%
  \BibitemOpen
  \bibfield  {author} {\bibinfo {author} {\bibfnamefont {T.}~\bibnamefont
  {Hatsuda}}\ and\ \bibinfo {author} {\bibfnamefont {T.}~\bibnamefont
  {Kunihiro}},\ }\href {\doibase 10.1016/0370-1573(94)90022-1} {\bibfield
  {journal} {\bibinfo  {journal} {Phys. Rept.}\ }\textbf {\bibinfo {volume}
  {247}},\ \bibinfo {pages} {221} (\bibinfo {year} {1994})},\ \Eprint
  {http://arxiv.org/abs/hep-ph/9401310} {arXiv:hep-ph/9401310 [hep-ph]}
  \BibitemShut {NoStop}%
%%CITATION = HEP-PH/9401310;%%
\bibitem [{\citenamefont {Neubert}(1994)}]{Neubert:1993mb}%
  \BibitemOpen
  \bibfield  {author} {\bibinfo {author} {\bibfnamefont {M.}~\bibnamefont
  {Neubert}},\ }\href {\doibase 10.1016/0370-1573(94)90091-4} {\bibfield
  {journal} {\bibinfo  {journal} {Phys. Rept.}\ }\textbf {\bibinfo {volume}
  {245}},\ \bibinfo {pages} {259} (\bibinfo {year} {1994})},\ \Eprint
  {http://arxiv.org/abs/hep-ph/9306320} {arXiv:hep-ph/9306320 [hep-ph]}
  \BibitemShut {NoStop}%
%%CITATION = HEP-PH/9306320;%%
\bibitem [{\citenamefont {Manohar}\ and\ \citenamefont
  {Wise}(2000)}]{Manohar:2000dt}%
  \BibitemOpen
  \bibfield  {author} {\bibinfo {author} {\bibfnamefont {A.~V.}\ \bibnamefont
  {Manohar}}\ and\ \bibinfo {author} {\bibfnamefont {M.~B.}\ \bibnamefont
  {Wise}},\ }\href@noop {} {\bibfield  {journal} {\bibinfo  {journal} {Camb.
  Monogr. Part. Phys. Nucl. Phys. Cosmol.}\ }\textbf {\bibinfo {volume} {10}},\
  \bibinfo {pages} {1} (\bibinfo {year} {2000})}\BibitemShut {NoStop}%
%%CITATION = CMPCE,10,1;%%
\bibitem [{\citenamefont {Yasui}\ \emph {et~al.}(2017)\citenamefont {Yasui},
  \citenamefont {Suzuki},\ and\ \citenamefont {Itakura}}]{Yasui:2017izi}%
  \BibitemOpen
  \bibfield  {author} {\bibinfo {author} {\bibfnamefont {S.}~\bibnamefont
  {Yasui}}, \bibinfo {author} {\bibfnamefont {K.}~\bibnamefont {Suzuki}}, \
  and\ \bibinfo {author} {\bibfnamefont {K.}~\bibnamefont {Itakura}},\ }\href
  {\doibase 10.1103/PhysRevD.96.014016} {\bibfield  {journal} {\bibinfo
  {journal} {Phys. Rev.}\ }\textbf {\bibinfo {volume} {D96}},\ \bibinfo {pages}
  {014016} (\bibinfo {year} {2017})},\ \Eprint
  {http://arxiv.org/abs/1703.04124} {arXiv:1703.04124 [hep-ph]} \BibitemShut
  {NoStop}%
%%CITATION = ARXIV:1703.04124;%%
\bibitem [{\citenamefont {Suzuki}\ \emph {et~al.}(2017)\citenamefont {Suzuki},
  \citenamefont {Yasui},\ and\ \citenamefont {Itakura}}]{Suzuki:2017gde}%
  \BibitemOpen
  \bibfield  {author} {\bibinfo {author} {\bibfnamefont {K.}~\bibnamefont
  {Suzuki}}, \bibinfo {author} {\bibfnamefont {S.}~\bibnamefont {Yasui}}, \
  and\ \bibinfo {author} {\bibfnamefont {K.}~\bibnamefont {Itakura}},\ }\href
  {\doibase 10.1103/PhysRevD.96.114007} {\bibfield  {journal} {\bibinfo
  {journal} {Phys. Rev.}\ }\textbf {\bibinfo {volume} {D96}},\ \bibinfo {pages}
  {114007} (\bibinfo {year} {2017})},\ \Eprint
  {http://arxiv.org/abs/1708.06930} {arXiv:1708.06930 [hep-ph]} \BibitemShut
  {NoStop}%
%%CITATION = ARXIV:1708.06930;%%
\bibitem [{\citenamefont {Buballa}(2005)}]{Buballa:2003qv}%
  \BibitemOpen
  \bibfield  {author} {\bibinfo {author} {\bibfnamefont {M.}~\bibnamefont
  {Buballa}},\ }\href {\doibase 10.1016/j.physrep.2004.11.004} {\bibfield
  {journal} {\bibinfo  {journal} {Phys. Rept.}\ }\textbf {\bibinfo {volume}
  {407}},\ \bibinfo {pages} {205} (\bibinfo {year} {2005})},\ \Eprint
  {http://arxiv.org/abs/hep-ph/0402234} {arXiv:hep-ph/0402234 [hep-ph]}
  \BibitemShut {NoStop}%
%%CITATION = HEP-PH/0402234;%%
\bibitem [{\citenamefont {Kanazawa}\ and\ \citenamefont
  {Uchino}(2016)}]{Kanazawa:2016ihl}%
  \BibitemOpen
  \bibfield  {author} {\bibinfo {author} {\bibfnamefont {T.}~\bibnamefont
  {Kanazawa}}\ and\ \bibinfo {author} {\bibfnamefont {S.}~\bibnamefont
  {Uchino}},\ }\href {\doibase 10.1103/PhysRevD.94.114005} {\bibfield
  {journal} {\bibinfo  {journal} {Phys. Rev.}\ }\textbf {\bibinfo {volume}
  {D94}},\ \bibinfo {pages} {114005} (\bibinfo {year} {2016})},\ \Eprint
  {http://arxiv.org/abs/1609.00033} {arXiv:1609.00033 [cond-mat.str-el]}
  \BibitemShut {NoStop}%
%%CITATION = ARXIV:1609.00033;%%
\bibitem [{\citenamefont {Yasui}(2017)}]{Yasui:2016yet}%
  \BibitemOpen
  \bibfield  {author} {\bibinfo {author} {\bibfnamefont {S.}~\bibnamefont
  {Yasui}},\ }\href {\doibase 10.1016/j.physletb.2017.08.066} {\bibfield
  {journal} {\bibinfo  {journal} {Phys. Lett.}\ }\textbf {\bibinfo {volume}
  {B773}},\ \bibinfo {pages} {428} (\bibinfo {year} {2017})},\ \Eprint
  {http://arxiv.org/abs/1608.06450} {arXiv:1608.06450 [hep-ph]} \BibitemShut
  {NoStop}%
%%CITATION = ARXIV:1608.06450;%%
\bibitem [{\citenamefont {Wilson}(1975)}]{Wilson:1974mb}%
  \BibitemOpen
  \bibfield  {author} {\bibinfo {author} {\bibfnamefont {K.~G.}\ \bibnamefont
  {Wilson}},\ }\href {\doibase 10.1103/RevModPhys.47.773} {\bibfield  {journal}
  {\bibinfo  {journal} {Rev. Mod. Phys.}\ }\textbf {\bibinfo {volume} {47}},\
  \bibinfo {pages} {773} (\bibinfo {year} {1975})}\BibitemShut {NoStop}%
%%CITATION = RMPHA,47,773;%%
\bibitem [{\citenamefont {Kojo}\ \emph {et~al.}(2010)\citenamefont {Kojo},
  \citenamefont {Hidaka}, \citenamefont {McLerran},\ and\ \citenamefont
  {Pisarski}}]{Kojo:2009ha}%
  \BibitemOpen
  \bibfield  {author} {\bibinfo {author} {\bibfnamefont {T.}~\bibnamefont
  {Kojo}}, \bibinfo {author} {\bibfnamefont {Y.}~\bibnamefont {Hidaka}},
  \bibinfo {author} {\bibfnamefont {L.}~\bibnamefont {McLerran}}, \ and\
  \bibinfo {author} {\bibfnamefont {R.~D.}\ \bibnamefont {Pisarski}},\ }\href
  {\doibase 10.1016/j.nuclphysa.2010.05.053} {\bibfield  {journal} {\bibinfo
  {journal} {Nucl. Phys.}\ }\textbf {\bibinfo {volume} {A843}},\ \bibinfo
  {pages} {37} (\bibinfo {year} {2010})},\ \Eprint
  {http://arxiv.org/abs/0912.3800} {arXiv:0912.3800 [hep-ph]} \BibitemShut
  {NoStop}%
%%CITATION = ARXIV:0912.3800;%%
\bibitem [{\citenamefont {Aono}(2013)}]{AONO:2013}%
  \BibitemOpen
  \bibfield  {author} {\bibinfo {author} {\bibfnamefont {T.}~\bibnamefont
  {Aono}},\ }\href {\doibase 10.7566/JPSJ.82.083703} {\bibfield  {journal}
  {\bibinfo  {journal} {J. Phys. Soc. Jpn.}\ }\textbf {\bibinfo {volume}
  {82}},\ \bibinfo {pages} {083703} (\bibinfo {year} {2013})},\ \Eprint
  {http://arxiv.org/abs/1208.6370} {arXiv:1208.6370 [cond-mat.mes-hall]}
  \BibitemShut {NoStop}%
\bibitem [{\citenamefont {Novoselov}\ \emph {et~al.}(2005)\citenamefont
  {Novoselov}, \citenamefont {Geim}, \citenamefont {Morozov}, \citenamefont
  {Jiang}, \citenamefont {Katsnelson}, \citenamefont {Grigorieva},
  \citenamefont {Dubonos},\ and\ \citenamefont {Firsov}}]{Novoselov:2005}%
  \BibitemOpen
  \bibfield  {author} {\bibinfo {author} {\bibfnamefont {K.~S.}\ \bibnamefont
  {Novoselov}}, \bibinfo {author} {\bibfnamefont {A.~K.}\ \bibnamefont {Geim}},
  \bibinfo {author} {\bibfnamefont {S.~V.}\ \bibnamefont {Morozov}}, \bibinfo
  {author} {\bibfnamefont {D.}~\bibnamefont {Jiang}}, \bibinfo {author}
  {\bibfnamefont {M.~I.}\ \bibnamefont {Katsnelson}}, \bibinfo {author}
  {\bibfnamefont {I.~V.}\ \bibnamefont {Grigorieva}}, \bibinfo {author}
  {\bibfnamefont {S.~V.}\ \bibnamefont {Dubonos}}, \ and\ \bibinfo {author}
  {\bibfnamefont {A.~A.}\ \bibnamefont {Firsov}},\ }\href {\doibase
  10.1038/nature04233} {\bibfield  {journal} {\bibinfo  {journal} {Nature}\
  }\textbf {\bibinfo {volume} {438}},\ \bibinfo {pages} {197} (\bibinfo {year}
  {2005})},\ \Eprint {http://arxiv.org/abs/cond-mat/0509330}
  {arXiv:cond-mat/0509330 [cond-mat.mes-hall]} \BibitemShut {NoStop}%
\bibitem [{\citenamefont {Zhang}\ \emph {et~al.}(2005)\citenamefont {Zhang},
  \citenamefont {Tan}, \citenamefont {Stormer},\ and\ \citenamefont
  {Kim}}]{Zhang:2005}%
  \BibitemOpen
  \bibfield  {author} {\bibinfo {author} {\bibfnamefont {Y.}~\bibnamefont
  {Zhang}}, \bibinfo {author} {\bibfnamefont {Y.-W.}\ \bibnamefont {Tan}},
  \bibinfo {author} {\bibfnamefont {H.~L.}\ \bibnamefont {Stormer}}, \ and\
  \bibinfo {author} {\bibfnamefont {P.}~\bibnamefont {Kim}},\ }\href {\doibase
  10.1038/nature04235} {\bibfield  {journal} {\bibinfo  {journal} {Nature}\
  }\textbf {\bibinfo {volume} {438}},\ \bibinfo {pages} {201} (\bibinfo {year}
  {2005})},\ \Eprint {http://arxiv.org/abs/cond-mat/0509355}
  {arXiv:cond-mat/0509355 [cond-mat.mes-hall]} \BibitemShut {NoStop}%
\bibitem [{\citenamefont {Hsieh}\ \emph {et~al.}(2008)\citenamefont {Hsieh},
  \citenamefont {Qian}, \citenamefont {Wray}, \citenamefont {Xia},
  \citenamefont {Hor}, \citenamefont {Cava},\ and\ \citenamefont
  {Hasan}}]{Hsieh:2008}%
  \BibitemOpen
  \bibfield  {author} {\bibinfo {author} {\bibfnamefont {D.}~\bibnamefont
  {Hsieh}}, \bibinfo {author} {\bibfnamefont {D.}~\bibnamefont {Qian}},
  \bibinfo {author} {\bibfnamefont {L.}~\bibnamefont {Wray}}, \bibinfo {author}
  {\bibfnamefont {Y.}~\bibnamefont {Xia}}, \bibinfo {author} {\bibfnamefont
  {Y.~S.}\ \bibnamefont {Hor}}, \bibinfo {author} {\bibfnamefont {R.~J.}\
  \bibnamefont {Cava}}, \ and\ \bibinfo {author} {\bibfnamefont {M.~Z.}\
  \bibnamefont {Hasan}},\ }\href {\doibase 10.1038/nature06843} {\bibfield
  {journal} {\bibinfo  {journal} {Nature}\ }\textbf {\bibinfo {volume} {452}},\
  \bibinfo {pages} {970} (\bibinfo {year} {2008})},\ \Eprint
  {http://arxiv.org/abs/0910.2420} {arXiv:0910.2420 [cond-mat.mes-hall]}
  \BibitemShut {NoStop}%
\bibitem [{\citenamefont {Mitchell}\ \emph {et~al.}(2013)\citenamefont
  {Mitchell}, \citenamefont {Schuricht}, \citenamefont {Vojta},\ and\
  \citenamefont {Fritz}}]{Mitchell:2013}%
  \BibitemOpen
  \bibfield  {author} {\bibinfo {author} {\bibfnamefont {A.~K.}\ \bibnamefont
  {Mitchell}}, \bibinfo {author} {\bibfnamefont {D.}~\bibnamefont {Schuricht}},
  \bibinfo {author} {\bibfnamefont {M.}~\bibnamefont {Vojta}}, \ and\ \bibinfo
  {author} {\bibfnamefont {L.}~\bibnamefont {Fritz}},\ }\href {\doibase
  10.1103/PhysRevB.87.075430} {\bibfield  {journal} {\bibinfo  {journal} {Phys.
  Rev. B}\ }\textbf {\bibinfo {volume} {87}},\ \bibinfo {pages} {075430}
  (\bibinfo {year} {2013})},\ \Eprint {http://arxiv.org/abs/1211.0034}
  {arXiv:1211.0034 [cond-mat.mes-hall]} \BibitemShut {NoStop}%
\bibitem [{\citenamefont {Kanao}\ \emph {et~al.}(2012)\citenamefont {Kanao},
  \citenamefont {Matsuura},\ and\ \citenamefont {Ogata}}]{KANAO:2012}%
  \BibitemOpen
  \bibfield  {author} {\bibinfo {author} {\bibfnamefont {T.}~\bibnamefont
  {Kanao}}, \bibinfo {author} {\bibfnamefont {H.}~\bibnamefont {Matsuura}}, \
  and\ \bibinfo {author} {\bibfnamefont {M.}~\bibnamefont {Ogata}},\ }\href
  {\doibase 10.1143/JPSJ.81.063709} {\bibfield  {journal} {\bibinfo  {journal}
  {J. Phys. Soc. Jpn.}\ }\textbf {\bibinfo {volume} {81}},\ \bibinfo {pages}
  {063709} (\bibinfo {year} {2012})},\ \Eprint {http://arxiv.org/abs/1205.2148}
  {arXiv:1205.2148 [cond-mat.mtrl-sci]} \BibitemShut {NoStop}%
\bibitem [{\citenamefont {Be\'ri}\ and\ \citenamefont
  {Cooper}(2012)}]{Beri:2012}%
  \BibitemOpen
  \bibfield  {author} {\bibinfo {author} {\bibfnamefont {B.}~\bibnamefont
  {Be\'ri}}\ and\ \bibinfo {author} {\bibfnamefont {N.~R.}\ \bibnamefont
  {Cooper}},\ }\href {\doibase 10.1103/PhysRevLett.109.156803} {\bibfield
  {journal} {\bibinfo  {journal} {Phys. Rev. Lett.}\ }\textbf {\bibinfo
  {volume} {109}},\ \bibinfo {pages} {156803} (\bibinfo {year} {2012})},\
  \Eprint {http://arxiv.org/abs/1206.2224} {arXiv:1206.2224
  [cond-mat.mes-hall]} \BibitemShut {NoStop}%
\bibitem [{\citenamefont {Altland}\ \emph {et~al.}(2014)\citenamefont
  {Altland}, \citenamefont {Be\'ri}, \citenamefont {Egger},\ and\ \citenamefont
  {Tsvelik}}]{Altland:2014}%
  \BibitemOpen
  \bibfield  {author} {\bibinfo {author} {\bibfnamefont {A.}~\bibnamefont
  {Altland}}, \bibinfo {author} {\bibfnamefont {B.}~\bibnamefont {Be\'ri}},
  \bibinfo {author} {\bibfnamefont {R.}~\bibnamefont {Egger}}, \ and\ \bibinfo
  {author} {\bibfnamefont {A.~M.}\ \bibnamefont {Tsvelik}},\ }\href {\doibase
  10.1088/1751-8113/47/26/265001} {\bibfield  {journal} {\bibinfo  {journal}
  {J. Phys. A: Math. Theor.}\ }\textbf {\bibinfo {volume} {47}},\ \bibinfo
  {pages} {265001} (\bibinfo {year} {2014})},\ \Eprint
  {http://arxiv.org/abs/1403.0113} {arXiv:1403.0113 [cond-mat.mes-hall]}
  \BibitemShut {NoStop}%
\bibitem [{\citenamefont {Cheng}\ \emph {et~al.}(2014)\citenamefont {Cheng},
  \citenamefont {Becker}, \citenamefont {Bauer},\ and\ \citenamefont
  {Lutchyn}}]{Cheng:2014}%
  \BibitemOpen
  \bibfield  {author} {\bibinfo {author} {\bibfnamefont {M.}~\bibnamefont
  {Cheng}}, \bibinfo {author} {\bibfnamefont {M.}~\bibnamefont {Becker}},
  \bibinfo {author} {\bibfnamefont {B.}~\bibnamefont {Bauer}}, \ and\ \bibinfo
  {author} {\bibfnamefont {R.~M.}\ \bibnamefont {Lutchyn}},\ }\href {\doibase
  10.1103/PhysRevX.4.031051} {\bibfield  {journal} {\bibinfo  {journal} {Phys.
  Rev. X}\ }\textbf {\bibinfo {volume} {4}},\ \bibinfo {pages} {031051}
  (\bibinfo {year} {2014})},\ \Eprint {http://arxiv.org/abs/1308.4156}
  {arXiv:1308.4156 [cond-mat.mes-hall]} \BibitemShut {NoStop}%
\bibitem [{\citenamefont {Eriksson}\ \emph {et~al.}(2014)\citenamefont
  {Eriksson}, \citenamefont {Mora}, \citenamefont {Zazunov},\ and\
  \citenamefont {Egger}}]{Eriksson:2014}%
  \BibitemOpen
  \bibfield  {author} {\bibinfo {author} {\bibfnamefont {E.}~\bibnamefont
  {Eriksson}}, \bibinfo {author} {\bibfnamefont {C.}~\bibnamefont {Mora}},
  \bibinfo {author} {\bibfnamefont {A.}~\bibnamefont {Zazunov}}, \ and\
  \bibinfo {author} {\bibfnamefont {R.}~\bibnamefont {Egger}},\ }\href
  {\doibase 10.1103/PhysRevLett.113.076404} {\bibfield  {journal} {\bibinfo
  {journal} {Phys. Rev. Lett.}\ }\textbf {\bibinfo {volume} {113}},\ \bibinfo
  {pages} {076404} (\bibinfo {year} {2014})},\ \Eprint
  {http://arxiv.org/abs/1404.5499} {arXiv:1404.5499 [cond-mat.mes-hall]}
  \BibitemShut {NoStop}%
\end{thebibliography}%

\end{document}